# Search for Gravitational Waves Associated with Gamma-Ray Bursts during the First Advanced LIGO Observing Run and Implications for the Origin of GRB 150906B

B. P. Abbott[1], R. Abbott[1], T. D. Abbott[2], M. R. Abernathy[3], F. Acernese[4,5], K. Ackley[6], C. Adams[7], T. Adams[8], P. Addesso[9],
R. X. Adhikari[1], V. B. Adya[10], C. Affeldt[10], M. Agathos[11], K. Agatsuma[11], N. Aggarwal[12], O. D. Aguiar[13], L. Aiello[14,15], A. Ain[16],
P. Ajith[17], B. Allen[10,18,19], A. Allocca[20,21], P. A. Altin[22], A. Ananyeva[1], S. B. Anderson[1], W. G. Anderson[18], S. Appert[1], K. Arai[1],
M. C. Araya[1], J. S. Areeda[23], N. Arnaud[24], K. G. Arun[25], S. Ascenzi[15,26], G. Ashton[10], M. Ast[27], S. M. Aston[7], P. Astone[28],
P. Aufmuth[19], C. Aulbert[10], A. Avila-Alvarez[23], S. Babak[29], P. Bacon[30], M. K. M. Bader[11], P. T. Baker[31], F. Baldaccini[32,33],
G. Ballardin[34], S. W. Ballmer[35], J. C. Barayoga[1], S. E. Barclay[36], B. C. Barish[1], D. Barker[37], F. Barone[4,5], B. Barr[36], L. Barsotti[12],
M. Barsuglia[30], D. Barta[38], J. Bartlett[37], I. Bartos[39], R. Bassiri[40], A. Basti[20,21], J. C. Batch[37], C. Baune[10], V. Bavigadda[34],
M. Bazzan[41,42], B. Bécsy[43], C. Beer[10], M. Bejger[44], I. Belahcene[24], M. Belgin[45], A. S. Bell[36], B. K. Berger[1], G. Bergmann[10],
C. P. L. Berry[46], D. Bersanetti[47,48], A. Bertolini[11], J. Betzwieser[7], S. Bhagwat[35], R. Bhandare[49], I. A. Bilenko[50], G. Billingsley[1],
C. R. Billman[6], J. Birch[7], R. Birney[51], O. Birnholtz[10], S. Biscans[1,12], A. Bisht[19], M. Bitossi[34], C. Biwer[35], M. A. Bizouard[24],
J. K. Blackburn[1], J. Blackman[52], C. D. Blair[53], D. G. Blair[53], R. M. Blair[37], S. Bloemen[54], O. Bock[10], M. Boer[55], G. Bogaert[55],
A. Bohe[29], F. Bondu[56], R. Bonnand[8], B. A. Boom[11], R. Bork[1], V. Boschi[20,21], S. Bose[16,57], Y. Bouffanais[30], A. Bozzi[34],
C. Bradaschia[21], P. R. Brady[18], V. B. Braginsky[50,144], M. Branchesi[58,59], J. E. Brau[60], T. Briant[61], A. Brillet[55], M. Brinkmann[10],
V. Brisson[24], P. Brockill[18], J. E. Broida[62], A. F. Brooks[1], D. A. Brown[35], D. D. Brown[46], N. M. Brown[12], S. Brunett[1],
C. C. Buchanan[2], A. Buikema[12], T. Bulik[63], H. J. Bulten[11,64], A. Buonanno[29,65], D. Buskulic[8], C. Buy[30], R. L. Byer[40], M. Cabero[10],
L. Cadonati[45], G. Cagnoli[66,67], C. Cahillane[1], J. Calderón Bustillo[45], T. A. Callister[1], E. Calloni[5,68], J. B. Camp[69], M. Canepa[47,48],
K. C. Cannon[70], H. Cao[71], J. Cao[72], C. D. Capano[10], E. Capocasa[30], F. Carbognani[34], S. Caride[73], J. Casanueva Diaz[24],
C. Casentini[15,26], S. Caudill[18], M. Cavaglià[74], F. Cavalier[24], R. Cavalieri[34], G. Cella[21], C. B. Cepeda[1], L. Cerboni Baiardi[58,59],
G. Cerretani[20,21], E. Cesarini[15,26], S. J. Chamberlin[75], M. Chan[36], S. Chao[76], P. Charlton[77], E. Chassande-Mottin[30],
B. D. Cheeseboro[31], H. Y. Chen[78], Y. Chen[52], H.-P. Cheng[6], A. Chincarini[48], A. Chiummo[34], T. Chmiel[79], H. S. Cho[80], M. Cho[65],
J. H. Chow[22], N. Christensen[62], Q. Chu[53], A. J. K. Chua[81], S. Chua[61], S. Chung[53], G. Ciani[6], F. Clara[37], J. A. Clark[45], F. Cleva[55],
C. Cocchieri[74], E. Coccia[14,15], P.-F. Cohadon[61], A. Colla[28,82], C. G. Collette[83], L. Cominsky[84], M. Constancio, Jr.[13], L. Conti[42],
S. J. Cooper[46], T. R. Corbitt[2], N. Cornish[85], A. Corsi[73], S. Cortese[34], C. A. Costa[13], M. W. Coughlin[62], S. B. Coughlin[86],
J.-P. Coulon[55], S. T. Countryman[39], P. Couvares[1], P. B. Covas[87], E. E. Cowan[45], D. M. Coward[53], M. J. Cowart[7], D. C. Coyne[1],
R. Coyne[73], J. D. E. Creighton[18], T. D. Creighton[88], J. Cripe[2], S. G. Crowder[89], T. J. Cullen[23], A. Cumming[36], L. Cunningham[36],
E. Cuoco[34], T. Dal Canton[69], G. Dálya[43], S. L. Danilishin[36], S. D'Antonio[15], K. Danzmann[10,19], A. Dasgupta[90],
C. F. Da Silva Costa[6], V. Dattilo[34], I. Dave[49], M. Davier[24], G. S. Davies[36], D. Davis[35], E. J. Daw[91], B. Day[45], R. Day[34], S. De[35],
D. DeBra[40], G. Debreczeni[38], J. Degallaix[66], M. De Laurentis[5,68], S. Deléglise[61], W. Del Pozzo[46], T. Denker[10], T. Dent[10],
V. Dergachev[29], R. De Rosa[5,68], R. T. DeRosa[7], R. DeSalvo[9], J. Devenson[51], R. C. Devine[31], S. Dhurandhar[16], M. C. Díaz[88],
L. Di Fiore[5], M. Di Giovanni[92,93], T. Di Girolamo[5,68], A. Di Lieto[20,21], S. Di Pace[28,82], I. Di Palma[28,29,82], A. Di Virgilio[21],
Z. Doctor[78], V. Dolique[66], F. Donovan[12], K. L. Dooley[74], S. Doravari[10], I. Dorrington[94], R. Douglas[36], M. Dovale Álvarez[46],
T. P. Downes[18], M. Drago[10], R. W. P. Drever[1], J. C. Driggers[37], Z. Du[72], M. Ducrot[8], S. E. Dwyer[37], T. B. Edo[91], M. C. Edwards[62],
A. Effler[7], H.-B. Eggenstein[10], P. Ehrens[1], J. Eichholz[1], S. S. Eikenberry[6], R. A. Eisenstein[12], R. C. Essick[12], Z. Etienne[31], T. Etzel[1],
M. Evans[12], T. M. Evans[7], R. Everett[75], M. Factourovich[39], V. Fafone[14,15,26], H. Fair[35], S. Fairhurst[94], X. Fan[72], S. Farinon[48],
B. Farr[78], W. M. Farr[46], E. J. Fauchon-Jones[94], M. Favata[95], M. Fays[94], H. Fehrmann[10], M. M. Fejer[40], A. Fernández Galiana[12],
I. Ferrante[20,21], E. C. Ferreira[13], F. Ferrini[34], F. Fidecaro[20,21], I. Fiori[34], D. Fiorucci[30], R. P. Fisher[35], R. Flaminio[66,96],
M. Fletcher[36], H. Fong[97], S. S. Forsyth[45], J.-D. Fournier[55], S. Frasca[28,82], F. Frasconi[21], Z. Frei[43], A. Freise[46], R. Frey[60], V. Frey[24],
E. M. Fries[1], P. Fritschel[12], V. V. Frolov[7], P. Fulda[6,69], M. Fyffe[7], H. Gabbard[74], B. U. Gadre[16], S. M. Gaebel[46], J. R. Gair[36],
L. Gammaitoni[32], S. G. Gaonkar[16], F. Garufi[5,68], G. Gaur[98], V. Gayathri[99], N. Gehrels[69], G. Gemme[48], E. Genin[34], A. Gennai[21],
J. George[49], L. Gergely[100], V. Germain[8], S. Ghonge[17], Abhirup Ghosh[17], A. Ghosh[11], Archisman Ghosh[17], S. Ghosh[11,54],
J. A. Giaime[2,7], K. D. Giardina[7], A. Giazotto[21], K. Gill[101], A. Glaefke[36], E. Goetz[10], R. Goetz[6], L. Gondan[43], G. González[2],
J. M. Gonzalez Castro[20,21], A. Gopakumar[102], M. L. Gorodetsky[50], S. E. Gossan[1], M. Gosselin[34], R. Gouaty[8], A. Grado[5,103],
C. Graef[36], M. Granata[66], A. Grant[36], S. Gras[12], C. Gray[37], G. Greco[58,59], A. C. Green[46], P. Groot[54], H. Grote[10], S. Grunewald[29],
G. M. Guidi[58,59], X. Guo[72], A. Gupta[16], M. K. Gupta[90], K. E. Gushwa[1], E. K. Gustafson[1], R. Gustafson[104], J. J. Hacker[23],
B. R. Hall[57], E. D. Hall[1], G. Hammond[36], M. Haney[102], M. M. Hanke[10], J. Hanks[37], C. Hanna[75], J. Hanson[7], T. Hardwick[2],
J. Harms[58,59], G. M. Harry[3], I. W. Harry[29], M. J. Hart[36], M. T. Hartman[6], C.-J. Haster[46,97], K. Haughian[36], J. Healy[105],
A. Heidmann[61], M. C. Heintze[7], H. Heitmann[55], P. Hello[24], G. Hemming[34], M. Hendry[36], I. S. Heng[36], J. Hennig[36], J. Henry[105],
A. W. Heptonstall[1], M. Heurs[10,19], S. Hild[36], D. Hoak[34], D. Hofman[66], K. Holt[7], D. E. Holz[78], P. Hopkins[94], J. Hough[36],
E. A. Houston[36], E. J. Howell[53], Y. M. Hu[10], E. A. Huerta[106], D. Huet[24], B. Hughey[101], S. Husa[87], S. H. Huttner[36], T. Huynh-Dinh[7],
N. Indik[10], D. R. Ingram[37], R. Inta[73], H. N. Isa[36], J.-M. Isac[61], M. Isi[1], T. Isogai[12], B. R. Iyer[17], K. Izumi[37], T. Jacqmin[61], K. Jani[45],
P. Jaranowski[107], S. Jawahar[108], F. Jiménez-Forteza[87], W. W. Johnson[2], D. I. Jones[109], R. Jones[36], R. J. G. Jonker[11], L. Ju[53],
J. Junker[10], C. V. Kalaghatgi[94], V. Kalogera[86], S. Kandhasamy[74], G. Kang[80], J. B. Kanner[1], S. Karki[60], K. S. Karvinen[10],

M. Kasprzack[2], E. Katsavounidis[12], W. Katzman[7], S. Kaufer[19], T. Kaur[53], K. Kawabe[37], F. Kéfélian[55], D. Keitel[87], D. B. Kelley[35], R. Kennedy[91], J. S. Key[88], F. Y. Khalili[50], I. Khan[14], S. Khan[94], Z. Khan[90], E. A. Khazanov[110], N. Kijbunchoo[37], Chunglee Kim[111], J. C. Kim[112], Whansun Kim[113], W. Kim[71], Y.-M. Kim[111,114], S. J. Kimbrell[45], E. J. King[71], P. J. King[37], R. Kirchhoff[10], J. S. Kissel[37], B. Klein[86], L. Kleybolte[27], S. Klimenko[6], P. Koch[10], S. M. Koehlenbeck[10], S. Koley[11], V. Kondrashov[1], A. Kontos[12], M. Korobko[27], W. Z. Korth[1], I. Kowalska[63], D. B. Kozak[1], C. Krämer[10], V. Kringel[10], B. Krishnan[10], A. Królak[115,116], G. Kuehn[10], P. Kumar[97], R. Kumar[90], L. Kuo[76], A. Kutynia[115], B. D. Lackey[29,35], M. Landry[37], R. N. Lang[18], J. Lange[105], B. Lantz[40], R. K. Lanza[12], A. Lartaux-Vollard[24], P. D. Lasky[117], M. Laxen[7], A. Lazzarini[1], C. Lazzaro[42], P. Leaci[28,82], S. Leavey[36], E. O. Lebigot[30], C. H. Lee[114], H. K. Lee[118], H. M. Lee[111], K. Lee[36], J. Lehmann[10], A. Lenon[31], M. Leonardi[92,93], J. R. Leong[10], N. Leroy[24], N. Letendre[8], Y. Levin[117], T. G. F. Li[119], A. Libson[12], T. B. Littenberg[120], J. Liu[53], N. A. Lockerbie[108], A. L. Lombardi[45], L. T. London[94], J. E. Lord[35], M. Lorenzini[14,15], V. Loriette[121], M. Lormand[7], G. Losurdo[21], J. D. Lough[10,19], G. Lovelace[23], H. Lück[10,19], A. P. Lundgren[10], R. Lynch[12], Y. Ma[52], S. Macfoy[51], B. Machenschalk[10], M. MacInnis[12], D. M. Macleod[2], F. Magaña-Sandoval[35], E. Majorana[28], I. Maksimovic[121], V. Malvezzi[15,26], N. Man[55], V. Mandic[122], V. Mangano[36], G. L. Mansell[22], M. Manske[18], M. Mantovani[34], F. Marchesoni[33,123], F. Marion[8], S. Márka[39], Z. Márka[39], A. S. Markosyan[40], E. Maros[1], F. Martelli[58,59], L. Martellini[55], I. W. Martin[36], D. V. Martynov[12], K. Mason[12], A. Masserot[8], T. J. Massinger[1], M. Masso-Reid[36], S. Mastrogiovanni[28,82], F. Matichard[1,12], L. Matone[39], N. Mavalvala[12], N. Mazumder[57], R. McCarthy[37], D. E. McClelland[22], S. McCormick[7], C. McGrath[18], S. C. McGuire[124], G. McIntyre[1], J. McIver[1], D. J. McManus[22], T. McRae[22], S. T. McWilliams[31], D. Meacher[55,75], G. D. Meadors[29], J. Meidam[11], A. Melatos[125], G. Mendell[37], D. Mendoza-Gandara[10], R. A. Mercer[18], E. L. Merilh[37], M. Merzougui[55], S. Meshkov[1], C. Messenger[36], C. Messick[75], R. Metzdorff[61], P. M. Meyers[122], F. Mezzani[28,82], H. Miao[46], C. Michel[66], H. Middleton[46], E. E. Mikhailov[126], L. Milano[5,68], A. L. Miller[6,28,82], A. Miller[86], B. B. Miller[86], J. Miller[12], M. Millhouse[85], Y. Minenkov[15], J. Ming[29], S. Mirshekari[127], C. Mishra[17], S. Mitra[16], V. P. Mitrofanov[50], G. Mitselmakher[6], R. Mittleman[12], A. Moggi[21], M. Mohan[34], S. R. P. Mohapatra[12], M. Montani[58,59], B. C. Moore[95], C. J. Moore[81], D. Moraru[37], G. Moreno[37], S. R. Morriss[88], B. Mours[8], C. M. Mow-Lowry[46], G. Mueller[6], A. W. Muir[94], Arunava Mukherjee[17], D. Mukherjee[18], S. Mukherjee[88], N. Mukund[16], A. Mullavey[7], J. Munch[71], E. A. M. Muniz[23], P. G. Murray[36], A. Mytidis[6], K. Napier[45], I. Nardecchia[15,26], L. Naticchioni[28,82], G. Nelemans[11,54], T. J. N. Nelson[7], M. Neri[47,48], M. Nery[10], A. Neunzert[104], J. M. Newport[3], G. Newton[36], T. T. Nguyen[22], A. B. Nielsen[10], S. Nissanke[11,54], A. Nitz[10], A. Noack[10], F. Nocera[34], D. Nolting[7], M. E. N. Normandin[88], L. K. Nuttall[35], J. Oberling[37], E. Ochsner[18], E. Oelker[12], G. H. Ogin[128], J. J. Oh[113], S. H. Oh[113], F. Ohme[10,94], M. Oliver[87], P. Oppermann[10], Richard J. Oram[7], B. O'Reilly[7], R. O'Shaughnessy[105], D. J. Ottaway[71], H. Overmier[7], B. J. Owen[73], A. E. Pace[75], J. Page[120], A. Pai[99], S. A. Pai[49], J. R. Palamos[60], O. Palashov[110], C. Palomba[28], A. Pal-Singh[27], H. Pan[76], C. Pankow[86], F. Pannarale[94], B. C. Pant[49], F. Paoletti[21,34], A. Paoli[34], M. A. Papa[10,18,29], H. R. Paris[40], W. Parker[7], D. Pascucci[36], A. Pasqualetti[34], R. Passaquieti[20,21], D. Passuello[21], B. Patricelli[20,21], B. L. Pearlstone[36], M. Pedraza[1], R. Pedurand[66,129], L. Pekowsky[35], A. Pele[7], S. Penn[130], C. J. Perez[37], A. Perreca[1], L. M. Perri[86], H. P. Pfeiffer[97], M. Phelps[36], O. J. Piccinni[28,82], M. Pichot[55], F. Piergiovanni[58,59], V. Pierro[9], G. Pillant[34], L. Pinard[66], I. M. Pinto[9], M. Pitkin[36], M. Poe[18], R. Poggiani[20,21], P. Popolizio[34], A. Post[10], J. Powell[36], J. Prasad[16], J. W. W. Pratt[101], V. Predoi[94], T. Prestegard[18,122], M. Prijatelj[10,34], M. Principe[9], S. Privitera[29], G. A. Prodi[92,93], L. G. Prokhorov[50], O. Puncken[10], M. Punturo[33], P. Puppo[28], M. Pürrer[29], H. Qi[18], J. Qin[53], S. Qiu[117], V. Quetschke[88], E. A. Quintero[1], R. Quitzow-James[60], F. J. Raab[37], D. S. Rabeling[22], H. Radkins[37], P. Raffai[43], S. Raja[49], C. Rajan[49], M. Rakhmanov[88], P. Rapagnani[28,82], V. Raymond[29], M. Razzano[20,21], V. Re[26], J. Read[23], T. Regimbau[55], L. Rei[48], S. Reid[51], D. H. Reitze[1,6], H. Rew[126], S. D. Reyes[35], E. Rhoades[101], F. Ricci[28,82], K. Riles[104], M. Rizzo[105], N. A. Robertson[1,36], R. Robie[36], F. Robinet[24], A. Rocchi[15], L. Rolland[8], J. G. Rollins[1], V. J. Roma[60], R. Romano[4,5], J. H. Romie[7], D. Rosińska[44,131], S. Rowan[36], A. Rüdiger[10], P. Ruggi[34], K. Ryan[37], S. Sachdev[1], T. Sadecki[37], L. Sadeghian[18], M. Sakellariadou[132], L. Salconi[34], M. Saleem[99], F. Salemi[10], A. Samajdar[133], L. Sammut[117], L. M. Sampson[86], E. J. Sanchez[1], V. Sandberg[37], J. R. Sanders[35], B. Sassolas[66], B. S. Sathyaprakash[75,94], P. R. Saulson[35], O. Sauter[104], R. L. Savage[37], A. Sawadsky[19], P. Schale[60], J. Scheuer[86], E. Schmidt[101], J. Schmidt[10], P. Schmidt[1,52], R. Schnabel[27], R. M. S. Schofield[60], A. Schönbeck[27], E. Schreiber[10], D. Schuette[10,19], S. G. Schwalbe[101], J. Scott[36], S. M. Scott[22], D. Sellers[7], A. S. Sengupta[134], D. Sentenac[34], V. Sequino[15,26], A. Sergeev[110], Y. Setyawati[11,54], D. A. Shaddock[22], T. J. Shaffer[37], M. S. Shahriar[86], B. Shapiro[40], P. Shawhan[65], A. Sheperd[18], D. H. Shoemaker[12], D. M. Shoemaker[45], K. Siellez[45], X. Siemens[18], M. Sieniawska[44], D. Sigg[37], A. D. Silva[13], A. Singer[1], L. P. Singer[69], A. Singh[10,19,29], R. Singh[2], A. Singhal[14], A. M. Sintes[87], B. J. J. Slagmolen[22], B. Smith[7], J. R. Smith[23], R. J. E. Smith[1], E. J. Son[113], B. Sorazu[36], F. Sorrentino[48], T. Souradeep[16], A. P. Spencer[36], A. K. Srivastava[90], A. Staley[39], M. Steinke[10], J. Steinlechner[36], S. Steinlechner[27,36], D. Steinmeyer[10,19], B. C. Stephens[18], S. P. Stevenson[46], R. Stone[88], K. A. Strain[36], N. Straniero[66], G. Stratta[58,59], S. E. Strigin[50], R. Sturani[127], A. L. Stuver[7], T. Z. Summerscales[135], L. Sun[125], S. Sunil[90], P. J. Sutton[94], B. L. Swinkels[34], M. J. Szczepańczyk[101], A. Szolgyen[43], M. Tacca[30], D. Talukder[60], D. B. Tanner[6], M. Tápai[100], A. Taracchini[29], R. Taylor[1], T. Theeg[10], E. G. Thomas[46], M. Thomas[7], P. Thomas[37], K. A. Thorne[7], E. Thrane[117], T. Tippens[45], S. Tiwari[14,93], V. Tiwari[94], K. V. Tokmakov[108], K. Toland[36], C. Tomlinson[91], M. Tonelli[20,21], Z. Tornasi[36], C. I. Torrie[1], D. Töyrä[46], F. Travasso[32,33], G. Traylor[7], D. Trifirò[74], J. Trinastic[6], M. C. Tringali[92,93], L. Trozzo[21,136], M. Tse[12], R. Tso[1], M. Turconi[55], D. Tuyenbayev[88], D. Ugolini[137], C. S. Unnikrishnan[102], A. L. Urban[1], S. A. Usman[94], H. Vahlbruch[19], G. Vajente[1], G. Valdes[88], N. van Bakel[11], M. van Beuzekom[11], J. F. J. van den Brand[11,64], C. Van Den Broeck[11], D. C. Vander-Hyde[35], L. van der Schaaf[11], J. V. van Heijningen[11], A. A. van Veggel[36], M. Vardaro[41,42], V. Varma[52], S. Vass[1], M. Vasúth[38], A. Vecchio[46], G. Vedovato[42], J. Veitch[46], P. J. Veitch[71], K. Venkateswara[138], G. Venugopalan[1], D. Verkindt[8],

F. Vetrano[58,59], A. Viceré[58,59], A. D. Viets[18], S. Vinciguerra[46], D. J. Vine[51], J.-Y. Vinet[55], S. Vitale[12], T. Vo[35], H. Vocca[32,33], C. Vorvick[37], D. V. Voss[6], W. D. Vousden[46], S. P. Vyatchanin[50], A. R. Wade[1], L. E. Wade[79], M. Wade[79], M. Walker[2], L. Wallace[1], S. Walsh[10,29], G. Wang[14,59], H. Wang[46], M. Wang[46], Y. Wang[53], R. L. Ward[22], J. Warner[37], M. Was[8], J. Watchi[83], B. Weaver[37], L.-W. Wei[55], M. Weinert[10], A. J. Weinstein[1], R. Weiss[12], L. Wen[53], P. Weßels[10], T. Westphal[10], K. Wette[10], J. T. Whelan[105], B. F. Whiting[6], C. Whittle[117], D. Williams[36], R. D. Williams[1], A. R. Williamson[94], J. L. Willis[139], B. Willke[10,19], M. H. Wimmer[10,19], W. Winkler[10], C. C. Wipf[1], H. Wittel[10,19], G. Woan[36], J. Woehler[10], J. Worden[37], J. L. Wright[36], D. S. Wu[10], G. Wu[7], W. Yam[12], H. Yamamoto[1], C. C. Yancey[65], M. J. Yap[22], Hang Yu[12], Haocun Yu[12], M. Yvert[8], A. Zadrożny[115], L. Zangrando[42], M. Zanolin[101], J.-P. Zendri[42], M. Zevin[86], L. Zhang[1], M. Zhang[126], T. Zhang[36], Y. Zhang[105], C. Zhao[53], M. Zhou[86], Z. Zhou[86], X. J. Zhu[53], M. E. Zucker[1,12], J. Zweizig[1]

(The LIGO Scientific Collaboration and the Virgo Collaboration),

and

R. L. Aptekar[140], D. D. Frederiks[140], S. V. Golenetskii[140], D. V. Golovin[141], K. Hurley[142], M. L. Litvak[141], I. G. Mitrofanov[141], A. Rau[143], A. B. Sanin[141], D. S. Svinkin[140], A. von Kienlin[143], X. Zhang[143]

(The IPN Collaboration)

[1] LIGO, California Institute of Technology, Pasadena, CA 91125, USA
[2] Louisiana State University, Baton Rouge, LA 70803, USA
[3] American University, Washington, DC 20016, USA
[4] Università di Salerno, Fisciano, I-84084 Salerno, Italy
[5] INFN, Sezione di Napoli, Complesso Universitario di Monte S.Angelo, I-80126 Napoli, Italy
[6] University of Florida, Gainesville, FL 32611, USA
[7] LIGO Livingston Observatory, Livingston, LA 70754, USA
[8] Laboratoire d'Annecy-le-Vieux de Physique des Particules (LAPP), Université Savoie Mont Blanc, CNRS/IN2P3, F-74941 Annecy-le-Vieux, France
[9] University of Sannio at Benevento, I-82100 Benevento, Italy and INFN, Sezione di Napoli, I-80100 Napoli, Italy
[10] Albert-Einstein-Institut, Max-Planck-Institut für Gravitationsphysik, D-30167 Hannover, Germany
[11] Nikhef, Science Park, 1098 XG Amsterdam, The Netherlands
[12] LIGO, Massachusetts Institute of Technology, Cambridge, MA 02139, USA
[13] Instituto Nacional de Pesquisas Espaciais, 12227-010 São José dos Campos, São Paulo, Brazil
[14] INFN, Gran Sasso Science Institute, I-67100 L'Aquila, Italy
[15] INFN, Sezione di Roma Tor Vergata, I-00133 Roma, Italy
[16] Inter-University Centre for Astronomy and Astrophysics, Pune 411007, India
[17] International Centre for Theoretical Sciences, Tata Institute of Fundamental Research, Bengaluru 560089, India
[18] University of Wisconsin-Milwaukee, Milwaukee, WI 53201, USA
[19] Leibniz Universität Hannover, D-30167 Hannover, Germany
[20] Università di Pisa, I-56127 Pisa, Italy
[21] INFN, Sezione di Pisa, I-56127 Pisa, Italy
[22] Australian National University, Canberra, Australian Capital Territory 0200, Australia
[23] California State University Fullerton, Fullerton, CA 92831, USA
[24] LAL, Univ. Paris-Sud, CNRS/IN2P3, Université Paris-Saclay, F-91898 Orsay, France
[25] Chennai Mathematical Institute, Chennai 603103, India
[26] Università di Roma Tor Vergata, I-00133 Roma, Italy
[27] Universität Hamburg, D-22761 Hamburg, Germany
[28] INFN, Sezione di Roma, I-00185 Roma, Italy
[29] Albert-Einstein-Institut, Max-Planck-Institut für Gravitationsphysik, D-14476 Potsdam-Golm, Germany
[30] APC, AstroParticule et Cosmologie, Université Paris Diderot, CNRS/IN2P3, CEA/Irfu, Observatoire de Paris, Sorbonne Paris Cité, F-75205 Paris Cedex 13, France
[31] West Virginia University, Morgantown, WV 26506, USA
[32] Università di Perugia, I-06123 Perugia, Italy
[33] INFN, Sezione di Perugia, I-06123 Perugia, Italy
[34] European Gravitational Observatory (EGO), I-56021 Cascina, Pisa, Italy
[35] Syracuse University, Syracuse, NY 13244, USA
[36] SUPA, University of Glasgow, Glasgow G12 8QQ, UK
[37] LIGO Hanford Observatory, Richland, WA 99352, USA
[38] Wigner RCP, RMKI, H-1121 Budapest, Konkoly Thege Miklós út 29-33, Hungary
[39] Columbia University, New York, NY 10027, USA
[40] Stanford University, Stanford, CA 94305, USA
[41] Università di Padova, Dipartimento di Fisica e Astronomia, I-35131 Padova, Italy
[42] INFN, Sezione di Padova, I-35131 Padova, Italy
[43] MTA Eötvös University, "Lendulet" Astrophysics Research Group, Budapest 1117, Hungary
[44] Nicolaus Copernicus Astronomical Center, Polish Academy of Sciences, 00-716, Warsaw, Poland
[45] Center for Relativistic Astrophysics and School of Physics, Georgia Institute of Technology, Atlanta, GA 30332, USA
[46] University of Birmingham, Birmingham B15 2TT, UK
[47] Università degli Studi di Genova, I-16146 Genova, Italy
[48] INFN, Sezione di Genova, I-16146 Genova, Italy
[49] RRCAT, Indore MP 452013, India
[50] Faculty of Physics, Lomonosov Moscow State University, Moscow 119991, Russia
[51] SUPA, University of the West of Scotland, Paisley PA1 2BE, UK
[52] Caltech CaRT, Pasadena, CA 91125, USA
[53] University of Western Australia, Crawley, Western Australia 6009, Australia
[54] Department of Astrophysics/IMAPP, Radboud University Nijmegen, P.O. Box 9010, 6500 GL Nijmegen, The Netherlands
[55] Artemis, Université Côte d'Azur, CNRS, Observatoire Côte d'Azur, CS 34229, F-06304 Nice Cedex 4, France
[56] Institut de Physique de Rennes, CNRS, Université de Rennes 1, F-35042 Rennes, France

[57] Washington State University, Pullman, WA 99164, USA
[58] Università degli Studi di Urbino "Carlo Bo," I-61029 Urbino, Italy
[59] INFN, Sezione di Firenze, I-50019 Sesto Fiorentino, Firenze, Italy
[60] University of Oregon, Eugene, OR 97403, USA
[61] Laboratoire Kastler Brossel, UPMC-Sorbonne Universités, CNRS, ENS-PSL Research University, Collège de France, F-75005 Paris, France
[62] Carleton College, Northfield, MN 55057, USA
[63] Astronomical Observatory Warsaw University, 00-478 Warsaw, Poland
[64] VU University Amsterdam, 1081 HV Amsterdam, The Netherlands
[65] University of Maryland, College Park, MD 20742, USA
[66] Laboratoire des Matériaux Avancés (LMA), CNRS/IN2P3, F-69622 Villeurbanne, France
[67] Université Claude Bernard Lyon 1, F-69622 Villeurbanne, France
[68] Università di Napoli "Federico II," Complesso Universitario di Monte S.Angelo, I-80126 Napoli, Italy
[69] NASA/Goddard Space Flight Center, Greenbelt, MD 20771, USA
[70] RESCEU, University of Tokyo, Tokyo, 113-0033, Japan
[71] University of Adelaide, Adelaide, South Australia 5005, Australia
[72] Tsinghua University, Beijing 100084, China
[73] Texas Tech University, Lubbock, TX 79409, USA
[74] The University of Mississippi, University, MS 38677, USA
[75] The Pennsylvania State University, University Park, PA 16802, USA
[76] National Tsing Hua University, Hsinchu City, 30013 Taiwan, Republic of China
[77] Charles Sturt University, Wagga Wagga, New South Wales 2678, Australia
[78] University of Chicago, Chicago, IL 60637, USA
[79] Kenyon College, Gambier, OH 43022, USA
[80] Korea Institute of Science and Technology Information, Daejeon 305-806, Korea
[81] University of Cambridge, Cambridge CB2 1TN, UK
[82] Università di Roma "La Sapienza," I-00185 Roma, Italy
[83] University of Brussels, Brussels B-1050, Belgium
[84] Sonoma State University, Rohnert Park, CA 94928, USA
[85] Montana State University, Bozeman, MT 59717, USA
[86] Center for Interdisciplinary Exploration & Research in Astrophysics (CIERA), Northwestern University, Evanston, IL 60208, USA
[87] Universitat de les Illes Balears, IAC3—IEEC, E-07122 Palma de Mallorca, Spain
[88] The University of Texas Rio Grande Valley, Brownsville, TX 78520, USA
[89] Bellevue College, Bellevue, WA 98007, USA
[90] Institute for Plasma Research, Bhat, Gandhinagar 382428, India
[91] The University of Sheffield, Sheffield S10 2TN, UK
[92] Università di Trento, Dipartimento di Fisica, I-38123 Povo, Trento, Italy
[93] INFN, Trento Institute for Fundamental Physics and Applications, I-38123 Povo, Trento, Italy
[94] Cardiff University, Cardiff CF24 3AA, UK
[95] Montclair State University, Montclair, NJ 07043, USA
[96] National Astronomical Observatory of Japan, 2-21-1 Osawa, Mitaka, Tokyo 181-8588, Japan
[97] Canadian Institute for Theoretical Astrophysics, University of Toronto, Toronto, Ontario M5S 3H8, Canada
[98] University and Institute of Advanced Research, Gandhinagar, Gujarat 382007, India
[99] IISER-TVM, CET Campus, Trivandrum Kerala 695016, India
[100] University of Szeged, Dóm tér 9, Szeged 6720, Hungary
[101] Embry-Riddle Aeronautical University, Prescott, AZ 86301, USA
[102] Tata Institute of Fundamental Research, Mumbai 400005, India
[103] INAF, Osservatorio Astronomico di Capodimonte, I-80131, Napoli, Italy
[104] University of Michigan, Ann Arbor, MI 48109, USA
[105] Rochester Institute of Technology, Rochester, NY 14623, USA
[106] NCSA, University of Illinois at Urbana-Champaign, Urbana, IL 61801, USA
[107] University of Białystok, 15-424 Białystok, Poland
[108] SUPA, University of Strathclyde, Glasgow G1 1XQ, UK
[109] University of Southampton, Southampton SO17 1BJ, UK
[110] Institute of Applied Physics, Nizhny Novgorod, 603950, Russia
[111] Seoul National University, Seoul 151-742, Korea
[112] Inje University Gimhae, 621-749 South Gyeongsang, Korea
[113] National Institute for Mathematical Sciences, Daejeon 305-390, Korea
[114] Pusan National University, Busan 609-735, Korea
[115] NCBJ, 05-400 Świerk-Otwock, Poland
[116] Institute of Mathematics, Polish Academy of Sciences, 00656 Warsaw, Poland
[117] Monash University, Victoria 3800, Australia
[118] Hanyang University, Seoul 133-791, Korea
[119] The Chinese University of Hong Kong, Shatin, NT, Hong Kong
[120] University of Alabama in Huntsville, Huntsville, AL 35899, USA
[121] ESPCI, CNRS, F-75005 Paris, France
[122] University of Minnesota, Minneapolis, MN 55455, USA
[123] Università di Camerino, Dipartimento di Fisica, I-62032 Camerino, Italy
[124] Southern University and A&M College, Baton Rouge, LA 70813, USA
[125] The University of Melbourne, Parkville, Victoria 3010, Australia
[126] College of William and Mary, Williamsburg, VA 23187, USA
[127] Instituto de Física Teórica, University Estadual Paulista/ICTP South American Institute for Fundamental Research, São Paulo SP 01140-070, Brazil
[128] Whitman College, 345 Boyer Avenue, Walla Walla, WA 99362, USA
[129] Université de Lyon, F-69361 Lyon, France
[130] Hobart and William Smith Colleges, Geneva, NY 14456, USA
[131] Janusz Gil Institute of Astronomy, University of Zielona Góra, 65-265 Zielona Góra, Poland
[132] King's College London, University of London, London WC2R 2LS, UK

[133] IISER-Kolkata, Mohanpur, West Bengal 741252, India
[134] Indian Institute of Technology, Gandhinagar Ahmedabad Gujarat 382424, India
[135] Andrews University, Berrien Springs, MI 49104, USA
[136] Università di Siena, I-53100 Siena, Italy
[137] Trinity University, San Antonio, TX 78212, USA
[138] University of Washington, Seattle, WA 98195, USA
[139] Abilene Christian University, Abilene, TX 79699, USA
[140] Ioffe Institute, Politekhnicheskaya 26, St. Petersburg 194021, Russia
[141] Space Research Institute, Russian Academy of Sciences, Moscow 117997, Russia
[142] University of California-Berkeley, Space Sciences Lab, 7 Gauss Way, Berkeley, CA 94720, USA
[143] Max-Planck-Institut für Extraterrestrische Physik, Giessenbachstraße 1, D-85748, Garching, Germany

## Abstract

We present the results of the search for gravitational waves (GWs) associated with $\gamma$-ray bursts detected during the first observing run of the Advanced Laser Interferometer Gravitational-Wave Observatory (LIGO). We find no evidence of a GW signal for any of the 41 $\gamma$-ray bursts for which LIGO data are available with sufficient duration. For all $\gamma$-ray bursts, we place lower bounds on the distance to the source using the optimistic assumption that GWs with an energy of $10^{-2}M_{\odot}c^2$ were emitted within the 16–500 Hz band, and we find a median 90% confidence limit of 71 Mpc at 150 Hz. For the subset of 19 short/hard $\gamma$-ray bursts, we place lower bounds on distance with a median 90% confidence limit of 90 Mpc for binary neutron star (BNS) coalescences, and 150 and 139 Mpc for neutron star–black hole coalescences with spins aligned to the orbital angular momentum and in a generic configuration, respectively. These are the highest distance limits ever achieved by GW searches. We also discuss in detail the results of the search for GWs associated with GRB 150906B, an event that was localized by the InterPlanetary Network near the local galaxy NGC 3313, which is at a luminosity distance of 54 Mpc ($z = 0.0124$). Assuming the $\gamma$-ray emission is beamed with a jet half-opening angle $\leqslant 30°$, we exclude a BNS and a neutron star–black hole in NGC 3313 as the progenitor of this event with confidence >99%. Further, we exclude such progenitors up to a distance of 102 Mpc and 170 Mpc, respectively.

## 1. Introduction

Gamma-ray bursts (GRBs) are among the most energetic astrophysical events observed in the electromagnetic spectrum. They are transient flashes of $\gamma$-radiation and are broadly classified as being *long* or *short*, depending on their duration and spectral hardness, mainly on the basis of data from the Burst and Transient Source Experiment on board the *Compton* Gamma-Ray *Observatory* (Nakar 2007; Berger 2014). Long GRBs have a duration that is greater than ~2 s and a softer spectrum; their origin is related to the core collapse of rapidly rotating massive stars (Woosley & Bloom 2006; Mösta et al. 2015), a hypothesis supported by observations of associated core-collapse supernovae (Hjorth & Bloom 2011). In this scenario, several (magneto)rotational instabilities may kick in and lead to the emission of gravitational waves (GWs; Modjaz 2011).

Short GRBs have a duration of less than ~2 s and a harder spectrum. Their progenitors are widely thought to be coalescing binary neutron star (BNS) or neutron star (NS)–black hole (BH) binary systems (Eichler et al. 1989; Paczynski 1991; Narayan et al. 1992; Lee & Ramirez-Ruiz 2007; Nakar 2007; Berger 2011), a hypothesis that was reinforced by the observation of a possible kilonova associated with GRB 130603B (Berger et al. 2013; Tanvir et al. 2013). Coalescing BNS and NS-BH binaries—collectively NS binaries—also produce a characteristic GW signal that is detectable by the current generation of interferometric GW detectors, such as the Advanced Laser Interferometer Gravitational-Wave Observatory (LIGO) and Virgo, up to distances of hundreds of megaparsecs (Abbott et al. 2016c). GW signals associated with this class of GRBs would provide new astrophysical insight into the progenitors of these transient phenomena. Specifically, an NS binary coalescence signal in coincidence with a short GRB would confirm the NS binary merger origin. In addition, it would allow us to measure the masses and spins of the binary components—possibly enabling us to distinguish between BNS and NS-BH progenitors (Kreidberg et al. 2012; Hannam et al. 2013) and to constrain the relative merger rates of these two classes of compact binaries—as well as to place constraints on the beaming angle and the NS equation of state (Chen & Holz 2013; Pannarale & Ohme 2014; Clark et al. 2015). We note that observations of nearby long GRBs without an accompanying supernova (Della Valle et al. 2006; Fynbo et al. 2006; Gal-Yam et al. 2006) and of short GRBs that exhibit an extended $\gamma$-ray emission that is softer than the prompt spike (Gehrels et al. 2006; Norris & Bonnell 2006; Norris et al. 2010, 2011; Sakamoto et al. 2011) may blur the divide between long and short GRBs of the standard, bimodal classification. On the basis of their properties and their host environments, van Putten et al. (2014) ascribe the origin of GRBs from both categories to compact binary mergers, as for canonical short GRBs. In the case of short GRBs with and without extended emission, other studies indicate that there is no evidence to distinguish between the two populations (Fong et al. 2013; Fong & Berger 2013).

[144] Deceased, March 2016.

The first Advanced LIGO Observing Run (O1) began on 2015 September 12 and continued until 2016 January 19. During the run, the two LIGO detectors (located in Hanford, WA and Livingston, LA) were operating with instrument noise 3–4 times lower than ever measured before in their most sensitive frequency band, [100, 300] Hz; at 50 Hz, the sensitivity improvement with respect to the initial LIGO detectors was a factor of ∼30 (for further details on detector performance, see Figure 1 in Abbott et al. 2016g, Figure 2 in Martynov et al. 2016, and discussions therein, as well as Abbott et al. 2016f and Nuttall et al. 2015). In the course of O1, the search for GWs emitted by binary BH systems yielded two unambiguously identified signals (Abbott et al. 2016j, 2016h) and a third possible signal (Abbott et al. 2016d). These successful results also sparked the first campaign to search for counterparts of Advanced LIGO sources, marking a milestone for transient astronomy and paving the way for multimessenger investigations of NS binary merger events in the years to come (Abbott et al. 2016i, 2016e).

In this paper, we present the results of a search for GWs associated with GRBs detected by the *Fermi* and *Swift* $\gamma$-ray satellites and by the InterPlanetary Network (IPN) during O1. From current observations, one expects most GRB events to be at distances that are too large for their associated GW signals to be detectable (the median redshift of the long and short GRB populations with established redshifts is ∼2 and ∼0.5, respectively; Berger 2014). However, a GRB in a data set under consideration may happen to fall within the range of GW detectors. For example, the smallest observed redshift to date of an optical GRB afterglow is $z = 0.0085$ ($\simeq$36 Mpc) for GRB 980425 (Galama et al. 1998; Kulkarni et al. 1998; Iwamoto et al. 1998; see Clark et al. 2015 for further details on the expected rate of joint short GRB and GW observations). The effort reported in this paper follows the one carried out with the initial LIGO and Virgo detectors, which found no evidence for GWs in coincidence with 508 GRBs detected between 2005 and 2010 (Aasi et al. 2014b). Three distinct searches were performed during O1: (1) a low-latency search to promptly identify coincidences in time between online GW searches and GRB events (Rapid VOEvent Coincidence Monitor or RAVEN; Urban 2016; see Section 4.1 for details), (2) a modeled search for NS binary mergers (`PyGRB`; Williamson et al. 2014; Nitz et al. 2016; see Section 4.2), (3) a search for generic (i.e., using minimal assumptions about the signal morphology), unmodeled GW transients (`X-Pipeline`; Sutton et al. 2010; see Section 4.3). We find no evidence of a GW signal associated with any of the GRBs in the sample, and we also rule out a collective signature of weak GW signals associated with the GRB population. We determine lower bounds on the distance to the progenitor of each GRB, and we constrain the fraction of observed GRB population at low redshifts.

Finally, we report on the specific case of the search for GWs associated with GRB 150906B (Golenetskii et al. 2015; Hurley et al. 2015). This event, detected by the IPN, was poorly placed for optical/infrared observations, but, as noted by Levan et al. (2015), the local galaxy NGC 3313 lies close to the GRB 150906B IPN error box, making it a viable host candidate for this event. Interestingly, NGC 3313 is at a luminosity distance of 54 Mpc and is therefore within the Advanced LIGO horizon for NS binary mergers.

## 2. GRB Sample

Our GRB sample contains events distributed through the Gamma-ray Coordinates Network (GCN) system,[145] supplemented by the *Swift*[146] (Lien et al. 2016) and *Fermi*[147] (Gruber et al. 2014; von Kienlin et al. 2014) trigger pages and the IPN (Hurley et al. 2003). Events distributed through the GCN system are ingested into the GW candidate event database (GraceDB)[148] within seconds of publication. The dedicated Vetting Automation and Literature Informed Database (VALID; Coyne 2015) cross-checks their time and localization parameters against the tables relative to each satellite and against the published catalog, and with automated literature searches.

In total there are 110 bursts recorded in the GCN and the IPN database during the period of interest (2015 September 12 to 2016 January 19). Twenty-three of them were detected solely by the IPN,[149] and about half of these were observed by a single spacecraft or two closely spaced ones and therefore could not be localized. We followed up all GRBs that occurred when at least one of the LIGO detectors was operating in a stable configuration. GW data segments that are flagged as being of poor quality are excluded from the analysis. The classification of GRBs into short and long is sometimes somewhat ambiguous. Our selection is based on the $T_{90}$ duration, which is the time interval over which 90% of the total background-subtracted photon counts are observed. A GRB is labeled short if its $T_{90} + T_{90,\mathrm{error}} < 2$ s. A GRB is labeled long if $T_{90} - T_{90,\mathrm{error}} > 4$ s. The remaining GRBs are labeled ambiguous. This separates the GRB sample into 23 short GRBs, 79 long GRBs, and 8 ambiguous GRBs.

Since binary mergers are particularly strong sources of GWs, we use the modeled search for NS binaries to analyze both short GRBs and ambiguous GRBs. This ensures that we include all short GRBs in the tail of the duration distribution. This search was able to analyze 19 events, which constitute ∼61% of the GRBs it could have targeted, had the GW detectors been operating with 100% duty cycle. This search can be run with data from one or more GW detectors (see Section 4.2), so the number is in line with the ∼61% and ∼52% duty cycles of the Hanford and the Livingston detectors, respectively. The generic unmodeled GW search is performed on all GRBs, regardless of their classification. In this case, results were obtained for 31 GRBs, that is, 31% of the events recorded during O1 with available sky location information. Keeping in mind that this search requires at least 660 s of data in coincidence from the two GW detectors (see Section 4.3), we note that the number is in line with the ∼40% duty cycle of the two Advanced LIGO detectors during O1. In total, with the two methods, we were able to process 41 GRB events, that is, 41% of the events recorded during O1 that had sky location information available. Eight of these events were analyzed in single-detector mode by the modeled search for NS binaries: the ability of this search to run with data from only one detector thus allows us to significantly increase our sample.

[145] GCN Circulars Archive: http://gcn.gsfc.nasa.gov/gcn3_archive.html.

[146] *Swift* GRB Archive: http://swift.gsfc.nasa.gov/archive/grb_table/. *Swift*/BAT Gamma-Ray Burst Catalog: http://swift.gsfc.nasa.gov/results/batgrbcat/.

[147] FERMIGBRST—*Fermi* GBM Burst Catalog: https://heasarc.gsfc.nasa.gov/W3Browse/fermi/fermigbrst.html.

[148] Moe, B., Stephens, B., and Brady, P., GraceDB—Gravitational Wave Candidate Event Database, https://gracedb.ligo.org/.

[149] Unlike the GCN sample, the IPN sample we describe is the subset of GRBs that took place during O1 for which at least one LIGO detector was operating. For this subset, a detailed IPN sky localization was performed.

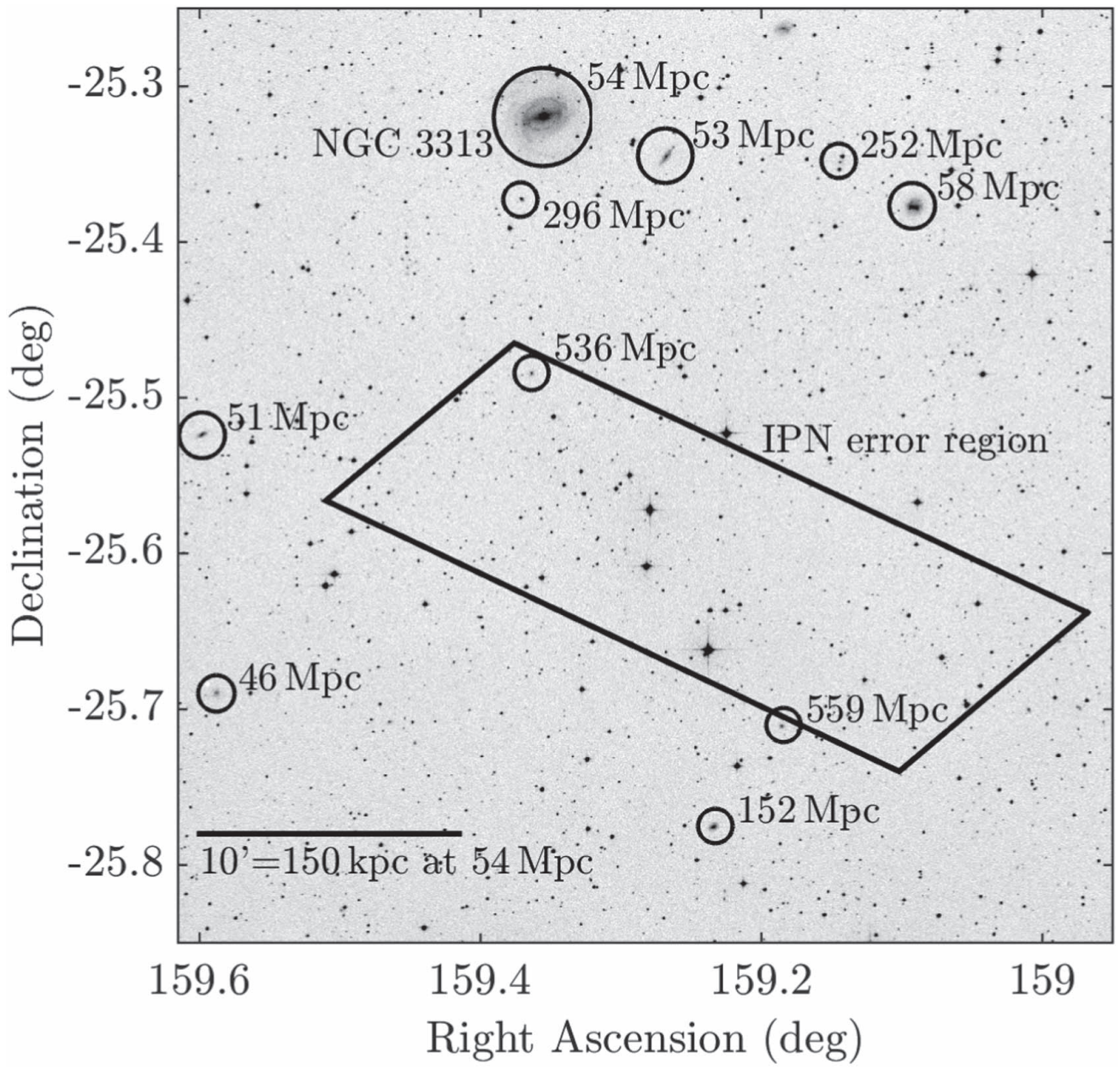

**Figure 1.** Overlay of the error box for GRB 150906B on the sky (Levan et al. 2015). A number of galaxies are at around 50 Mpc, while some of the galaxies within the error region are at ∼500 Mpc (G. Dálya et al. 2016, in preparation).

### 2.1. GRB 150906B

In addition to the GRBs in the sample we described above, we also consider GRB 150906B, an event of particular interest due to its potential proximity. It occurred on 2015 September 6 at 08:42:20.560 UTC and was detected by the IPN (Golenetskii et al. 2015; Hurley et al. 2015). At the time of GRB 150906B, the Advanced LIGO detectors were undergoing final preparations for O1. Nonetheless, the 4 km detector in Hanford was operational at that time.

GRB 150906B was observed by the *Konus–Wind*, *INTEGRAL*, *Mars Odyssey*, and *Swift* satellites. It was outside the coded field of view of the *Swift* BAT, and, consequently, localization was achieved by triangulation of the signals observed by the four satellites (Hurley et al. 2015). The localization region of GRB 150906B lies close to the local galaxy NGC 3313, which has a redshift of 0.0124 at a luminosity distance of 54 Mpc (Levan et al. 2015). This galaxy lies 130 kpc in projection from the GRB error box, a distance that is consistent with observed offsets of short GRBs from galaxies and with the expected supernova kicks imparted on NS binary systems (Berger 2011). NGC 3313 is part of a group of galaxies, and it is the brightest among this group. Other, fainter members of the group also lie close to the GRB error region, as shown in Figure 1. In addition, there are a number of known galaxies at around 500 Mpc within the error region of the GRB (Bilicki et al. 2013). For the GW search, we use a larger error region with a more conservative error assumption. Follow-up electromagnetic observations of the GRB were not possible due to its proximity to the Sun.

The *Konus–Wind* observation of GRB 150906B was further used to classify the GRB (Svinkin et al. 2015). It was observed to have a duration of[150] $T_{50} = (0.952 \pm 0.036)$ s and $T_{90} = (1.642 \pm 0.076)$ s, which places it at the longer end of the short GRB distribution. Furthermore, GRB 150906B lies between the peaks of the short/hard and long/soft *Konus–Wind* GRB distributions in the $\log T_{50}$–$\log \mathrm{HR}_{32}$ hardness–duration diagram, where $\log \mathrm{HR}_{32}$ is the (logarithm of the) ratio of counts in the [200, 760] keV and [50, 200] keV bands (Svinkin et al. 2015). Thus, a firm classification of the GRB as either short or long is problematic.

Assuming GRB 150906B originated in NGC 3313 yields an isotropic-equivalent $\gamma$-ray energy $E_{\mathrm{iso}} \sim 10^{49}$ erg (Levan et al. 2015). This is consistent with inferred luminosities of short GRBs with measured redshifts (Berger 2011), albeit at the lower end of the distribution of $E_{\mathrm{iso}}$ values. Theoretical arguments (Ruffini et al. 2015; Zhang et al. 2015) suggest that the energetics fit better with a more distant system around 500 Mpc, possibly originating from one of the galaxies within the error region.

## 3. Considerations on GRB Progenitors

As discussed previously, BNS and NS-BH mergers are the most plausible progenitors for the majority of short GRBs, while the progenitors of long GRBs are extreme cases of stellar collapse. In this section, we provide considerations on the main properties of the sources that we target with our searches in order to address these scenarios.

### 3.1. Short-duration GRBs

The modeled search for GWs emitted by NS binary mergers addresses the case of short GRB events. While not all NS binary mergers necessarily lead to a short GRB, this search looks for a GW counterpart to a short GRB event under the assumption that short GRBs are generated by NS binary mergers. In the standard scenario (Eichler et al. 1989; Paczynski 1991; Narayan et al. 1992; Nakar 2007), as the two companions spiral inward together due to the emission of GWs, the NSs are expected to tidally disrupt before the coalescence, in order to create a massive torus remnant in the surroundings of the central compact object that is formed by the binary coalescence. The matter in the torus can then power highly relativistic jets along the axis of total angular momentum (Blandford & Znajek 1977; Rosswog & Ramirez-Ruiz 2002; Lee & Ramirez-Ruiz 2007). This picture is supported by observational evidence (Berger 2011; Berger et al. 2013; Tanvir et al. 2013) and numerical simulations (e.g., Rezzolla et al. 2011; Kiuchi et al. 2015) but has not yet been fully confirmed.

The form of the GW signal emitted by a compact binary coalescence depends on the masses ($m_{\mathrm{NS}}$, $m_{\mathrm{comp}}$) and spins of the NS and its companion (either an NS or a BH), as well as the spatial location and orientation relative to the detector. In the remainder of this section we therefore discuss observational constraints on these properties and our choices regarding them that are folded into our search for BNS and NS-BH progenitors of short GRBs.

[150] Similarly to $T_{90}$, $T_{50}$ is the time interval over which 50% of the total background-subtracted photon counts are observed.

Mass measurements of NSs in binary systems currently set a lower bound on the maximum possible NS mass to $(2.01 \pm 0.04)\ M_\odot$ (Antoniadis et al. 2013). On the other hand, theoretical considerations set an upper bound on the maximum NS star mass to $\sim 3\ M_\odot$ (Rhoades & Ruffini 1974; Kalogera & Baym 1996), while the standard core-collapse supernova formation scenario restricts NS birth masses above the $1.1\text{–}1.6\ M_\odot$ interval (Lattimer 2012; Ozel et al. 2012; Kiziltan et al. 2013). Finally, we note that the individual NS masses reported for the eight candidate BNS systems lie in the interval $[1.0, 1.49]\ M_\odot$ (Ozel & Freire 2016).

The fastest spinning pulsar ever observed rotates at a frequency of 716 Hz (Hessels et al. 2006). Assuming a mass of $1.4\ M_\odot$ and a moment of inertia of $10^{45}$ g cm$^2$, this corresponds to a dimensionless spin magnitude of $\sim 0.4$. The highest measured spin frequency of pulsars in confirmed BNS systems is that of J0737−3039A (Burgay et al. 2003). It is equal to 44 Hz (Kramer & Wex 2009), which yields a dimensionless spin magnitude of $\sim 0.05$ (Brown et al. 2012). Finally, the potential BNS pulsar J1807−2500B (Lynch et al. 2012) with a spin of 4.19 ms gives a dimensionless spin magnitude of $\sim 0.2$, if one assumes a pulsar mass of $1.37\ M_\odot$ and a moment of inertia $2 \cdot 10^{45}$ g cm$^2$.

No observations of NS-BH systems are available to date. Notably, however, a likely NS-BH progenitor has been observed, namely Cyg X-3 (Belczynski et al. 2013). While Advanced LIGO has observed a BH with mass $36^{+5}_{-4}\ M_\odot$ in a binary BH system (Abbott et al. 2016j), and while stellar BHs with masses exceeding even $100\ M_\odot$ are conceivable (Belczynski et al. 2014; de Mink & Belczynski 2015), mass measurements of galactic stellar-mass BHs in X-ray binaries are between 5 and 24 solar masses (Ozel et al. 2010; Farr et al. 2011; Kreidberg et al. 2012; Wiktorowicz et al. 2013). X-ray observations of accreting BHs provide a broad distribution of dimensionless spin magnitudes ranging from $\sim 0.1$ to above 0.95 (e.g., Miller & Miller 2014). We remark that BH dimensionless spin magnitudes inferred from observations of high-mass X-ray binaries typically have values above 0.85 and that these systems are more likely to be NS-BH system progenitors (McClintock et al. 2014).

A final property to discuss in the context of GW searches for BNS and NS-BH systems in coincidence with short GRBs is the half-opening angle $\theta_{\rm jet}$ of the GRB jet. Relativistic beaming and collimation due to the ambient medium confine the GRB jet to $\theta_{\rm jet}$. In all cases, we assume that the GRB is emitted in the direction of the binary total angular momentum. The observation of prompt $\gamma$-ray emission is, therefore, indicative that the inclination of the total angular momentum with respect to the line of sight to the detectors lies within the jet cone. Estimates of $\theta_{\rm jet}$ are based on jet breaks observed in X-ray afterglows and vary across GRBs. Indeed, many GRBs do not even exhibit a jet break. However, studies of observed jet breaks in *Swift* GRB X-ray afterglows find a mean (median) value of $\theta_{\rm jet} = 6\overset{\circ}{.}5$ ($5\overset{\circ}{.}4$), with a tail extending almost to 25° (Racusin et al. 2009). In at least one case where no jet break is observed, the inferred lower limit is 25° and could be as high as 79° (Grupe et al. 2006). By folding in lower limits on $\theta_{\rm jet}$ for short GRBs without opening angle measurements and the indication that $\theta_{\rm jet} \sim 5°\text{–}20°$, which arises from simulations of postmerger BH accretion, Fong et al. (2015) find a median of $16° \pm 10°$ for $\theta_{\rm jet}$.

In light of all these considerations on astrophysical observations, we perform the modeled search described in Section 4.2 for NSs with masses between $1\ M_\odot$ and $2.8\ M_\odot$ and dimensionless spin magnitude of 0.05 at most.[151] For the companion object, we test masses in the range $1\ M_\odot \leqslant m_{\rm comp} \leqslant 25\ M_\odot$ and dimensionless spins up to 0.999. Additionally, we restrict the NS-BH search space (i.e., $m_{\rm comp} > 2.8\ M_\odot$) to BH masses and spins that are consistent with the presence of remnant material in the surroundings of the central BH, rather than with the direct plunge of the NS onto the BH (Pannarale & Ohme 2014). This astrophysically motivated cut excludes from our search NS-BH systems that do not allow for a GRB counterpart to be produced, even under the most optimistic assumptions regarding the NS equation of state[152] and the amount of tidally disrupted NS material required to ignite the GRB emission[153] (Pannarale & Ohme 2014). Finally, we search for circularly polarized signals. As discussed in Williamson et al. (2014), this is an excellent approximation for inclination angles between the total angular momentum and the line of sight up to 30°.

### 3.2. Long-duration GRBs

Long GRBs are followed up by the search for unmodeled GW transients described in Section 4.3. When making quantitative statements on the basis of this search, we use two families of GW signal models: circular sine-Gaussian (CSG) and accretion disk instability (ADI) signals. The scenarios that these address are discussed below.

No precise waveform is known for stellar collapse. A wide class of scenarios involves a rotational instability developing in the GRB central engine that leads to a slowly evolving, rotating quadrupolar mass distribution. Semianalytical calculations of rotational instabilities suggest that up to $10^{-2}\ M_\odot c^2$ may be emitted in GWs (Davies et al. 2002; Fryer et al. 2002; Kobayashi & Meszaros 2003; Shibata et al. 2003; Piro & Pfahl 2007; Corsi & Meszaros 2009; Romero et al. 2010), but simulations addressing the nonextreme case of core-collapse supernovae predict an emission of up to $10^{-8}\ M_\odot c^2$ in GWs (Ott 2009). With this in mind, we use a crude but simple generic model, that is, a CSG waveform with plus (+) and cross (×) polarizations given by

$$\begin{bmatrix} h_+(t) \\ h_\times(t) \end{bmatrix} = \frac{1}{r}\sqrt{\frac{G}{c^3}\frac{E_{\rm GW}}{f_0 Q}\frac{5}{4\pi^{3/2}}} \times \begin{bmatrix} (1+\cos^2\iota)\cos(2\pi f_0 t) \\ 2\cos\iota\sin(2\pi f_0 t) \end{bmatrix} \exp\left[-\frac{(2\pi f_0 t)^2}{2Q^2}\right], \quad (1)$$

where the signal frequency $f_0$ is equal to twice the rotation frequency, $t$ is the time relative to the signal peak time, $Q$

[151] The search is nonetheless effective for NS spins up to 0.4 (Nitz 2015; Abbott et al. 2016c).

[152] To prescribe the cut, we use a simple piecewise polytropic equation of state (2H) that yields NSs with masses up to $\sim 2.8\ M_\odot$ and radii of $\sim 15$ km (e.g., Kyutoku et al. 2010). The large NS radius value, which is above current constraints (Steiner et al. 2013; Ozel & Freire 2016), is chosen to favor tidal disruption and hence make our targeted parameter space as inclusive as possible.

[153] Namely, we target any system that leads to the presence of remnant NS debris material.

characterizes the number of cycles for which the quadrupolar mass moment is large, $E_{GW}$ is the total radiated energy, $r$ is the distance to the source, $\iota$ is the rotation axis inclination angle with respect to the observer, and $G$ and $c$ are the gravitational constant and the speed of light, respectively. The inclination angle $\iota$ can be once again linked to observations of GRB jet-opening angles: in the case of long GRBs, these are typically $\sim 5°$ (Gal-Yam et al. 2006; Racusin et al. 2009). All other parameters are largely underconstrained.

In the collapsar model of long GRBs, a stellar-mass BH forms, surrounded by a massive accretion disk. An extreme scenario of emission from a stellar collapse is a "magnetically suspended" ADI (van Putten 2001; van Putten et al. 2004). This parametric model may not be a precise representation of realistic signals, but it captures the generic features of many proposed models. It is characterized by four free astrophysical parameters: the mass and the dimensionless spin parameter of the central BH, the fraction of the disk mass that forms clumps, and the accretion disk mass. Waveform parameters such as duration, frequency span, and total radiated energy can be precisely derived for a given set of astrophysical parameters. As discussed in Section 4.3, we use several combinations of values for the astrophysical parameters in order to cover the different predicted morphologies.

## 4. Search Methodology

A *low-latency search*, referred to as RAVEN (Urban 2016; see Section 4.1), was performed in order to potentially initiate a prompt electromagnetic follow-up effort in the case of a possible association of a GW signal with a GRB event. The method builds on the results of the online, low-latency, all-sky GW searches to look for associations between GRBs and GW candidates. Results were ready within minutes after GRB detection notices: this allows circulation of alerts to the astronomy community on a timescale that is useful for enhancing follow-up observations targeting the X-ray, optical, and radio afterglows of GRB events.

The results presented in this paper were produced by an *offline search* using (1) a templated, NS binary search method (implemented in the `PyGRB` pipeline; see Williamson et al. 2014 and references therein for a full description and Nitz et al. 2016 for the code) for triggers corresponding to short GRBs and (2) a generic method (i.e., using minimal assumptions about the signal morphology) for GW transients (implemented in the `X-Pipeline`; see Sutton et al. 2010 and Was et al. 2012 for a complete description) for all GRBs. The two methods are illustrated in Sections 4.2 and 4.3, respectively. Both of them are largely the same as for the previous analysis described in Aasi et al. (2014b) and utilized data with final quality and calibration[154] (Abbott et al. 2017b, 2016f). Unlike in previous studies (Abbott et al. 2010; Abadie et al. 2010, 2012a; Aasi et al. 2014b), the offline search did not require data from both interferometers to be available. However, the generic method is severely limited by nonstationary transients when data from only one interferometer are available. Hence for the generic method we present results only for GRBs that occurred when both interferometers were available.

### 4.1. Rapid VOEvent Coincidence Monitor

RAVEN (Urban 2016) compares the GW triggers recorded in the low-latency all-sky GW analysis with the given time of a GRB. It provides a preliminary indication of any coincident GW candidate event and its associated significance. The `cWB` (Klimenko et al. 2016), `oLIB` (Lynch et al. 2015), `GstLAL` (Messick et al. 2017), and `MBTA` (Adams et al. 2016) pipelines perform the blind, rapid all-sky GW monitoring. `cWB` and `oLIB` search for a broad range of GW transients in the frequency range of 16–2048 Hz without prior knowledge of the signal waveforms. The `GstLAL` and `MBTA` pipelines search for GW signals from the coalescence of compact objects, using optimal matched filtering with waveforms. During O1, `MBTA` covered component masses of 1–12 $M_{\odot}$ with a 5 $M_{\odot}$ limit on chirp mass. `GstLAL`, instead, covered systems with component masses of 1–2.8 $M_{\odot}$ and 1–16 $M_{\odot}$ up to 2015 December 23; then, motivated by the discovery of GW150914, the analysis was extended to cover systems with component masses of 1–99 $M_{\odot}$ and total mass less than 100 $M_{\odot}$. Both pipelines limit component spins to $<0.99$ and $<0.05$ for BHs and NSs,[155] respectively (see Abbott et al. 2016c for further details).

GW candidates from these low-latency searches were uploaded to GraceDB and compared to the GRB triggers to find any temporal coincidence in $[-600, +60]$ and $[-5, +1]$ second windows, which correspond to the delay between the GW and the GRB trigger for long and short GRBs, respectively, as discussed in the next two sections. This strategy has the advantage of being very low latency and of requiring little additional computational costs over the existing all-sky searches. RAVEN (Urban 2016) results are available to be shared with LIGO partner electromagnetic astronomy facilities[156] within minutes following a GRB detection.

### 4.2. Neutron Star Binary Search Method (PyGRB)

In the vast majority of short GRB progenitor scenarios, the GW signal from an NS binary coalescence is expected to precede the prompt $\gamma$-ray emission by no more than a few seconds (Lee & Ramirez-Ruiz 2007; Vedrenne & Atteia 2009). Therefore, we search for NS binary GW signals with end times that lie in an *on-source* window of $[-5, +1)$ s around the reported GRB time, as done in previous searches in LIGO and Virgo data (Abadie et al. 2012a; Aasi et al. 2014b). The method we use is described in detail in Williamson et al. (2014) and references therein; the code implementing it is available under Nitz et al. (2016).

The data are filtered in the 30 Hz–1000 Hz frequency interval through a discrete bank of $\sim$110,000 template waveforms (Owen & Sathyaprakash 1999) that covers NS binaries with the properties discussed in Section 3.1. It is the first time that a short GRB follow-up search used a template bank that includes aligned spin systems (Brown et al. 2012; Harry et al. 2014). The bank is designed to have a 3%

[154] Both flavors of the search were also promptly initiated in a *medium-latency* configuration within about 20 minutes following the receipt of an appropriate GRB detection notice. This configuration requires a less accurate evaluation of the efficiency of each search and produces results within a few hours.

[155] `GstLAL` and `MBTA` treat as NSs components with masses below 2.8 $M_{\odot}$ and 2 $M_{\odot}$, respectively.

[156] See program description and participation information at http://www.ligo.org/scientists/GWEMalerts.php.

**Table 1**
Accretion Disk Instability Waveform Parameters

| Waveform Label | $M$ $(M_\odot)$ | $\chi$ | $\epsilon$ | Duration (s) | Frequency (Hz) | $E_{\rm GW}$ $(M_\odot c^2)$ |
|---|---|---|---|---|---|---|
| ADI-A | 5 | 0.30 | 0.050 | 39 | 135–166 | 0.02 |
| ADI-B | 10 | 0.95 | 0.200 | 9 | 110–209 | 0.22 |
| ADI-C | 10 | 0.95 | 0.040 | 236 | 130–251 | 0.25 |
| ADI-D | 3 | 0.70 | 0.035 | 142 | 119–173 | 0.02 |
| ADI-E | 8 | 0.99 | 0.065 | 76 | 111–234 | 0.17 |

**Notes.** The first column is the label used for the ADI waveform. The second and third columns are the mass and the dimensionless spin parameter of the central BH. The fourth column, $\epsilon$, is the fraction of the disk mass that forms clumps, and in all cases the accretion disk mass is $1.5\,M_\odot$. The duration, frequency span, and total radiated energy of the resulting signal are also reported in the remaining columns.

maximum loss of signal-to-noise ratio (S/N) due to discretization effects for binaries with spins aligned, or antialigned, to the orbital angular momentum over the parameter space discussed at the end of Section 3.1.

For BNS and NS-BH coalescences, we use point-particle post-Newtonian models that describe the inspiral stage, where the orbit of the binary slowly shrinks due to the emission of GWs. This is mainly motivated by the fact that the merger and postmerger regime (i.e., the GW high-frequency behavior) of these systems differs from the binary BH case. While we do have robust inspiral-merger-ringdown binary BH waveforms (Taracchini et al. 2014; Khan et al. 2016), efforts to obtain accurate, complete waveform models for NS binaries are still underway (Lackey et al. 2014; Bernuzzi et al. 2015; Pannarale et al. 2015; Barkett et al. 2016; Haas et al. 2016; Hinderer et al. 2016). Additionally, to go beyond a point-particle inspiral description, the search would have to cover all feasible NS equations of state, at the expense of a significant increase in its computational costs. Each template is therefore modeled with the "TaylorT4" time-domain, post-Newtonian inspiral approximant (Buonanno et al. 2003), filtered against the coherently combined data, and peaks in the matched filter coherent S/N are recorded. Additional signal consistency tests are used to eliminate the effect of non-Gaussian transients in the data and to generate a reweighted coherent S/N (see Williamson et al. 2014 for its formal definition), which forms the detection statistic (Allen 2005; Harry & Fairhurst 2011).

After the filtering and the consistency tests, the event with the largest reweighted coherent S/N (provided that this is greater than 6) in the on-source window is retained as a candidate GW signal. In order to assess the significance of the candidate, the detector background noise distribution is estimated using data from a time period surrounding the on-source data, when a GW signal is not expected to be present. This is known as the *off-source* data and is processed identically to the on-source data. Specifically, the same data-quality cuts and consistency tests are applied, and the same sky positions relative to the GW detector network are used. The NS binary search method requires a minimum of 1658 s of off-source data, which it splits into as many 6 s trials as possible. In order to increase the number of background trials, when data from more than one detector are available, the data streams are time-shifted multiple times and reanalyzed for each time shift. The template that produces the largest reweighted coherent S/N in each 6 s off-source time window is retained as a trigger. These are used to calculate a $p$-value[157] to the on-source loudest event by comparing it to the distribution of loudest off-source triggers in terms of the detection statistics. The $p$-value is calculated by counting the fraction of background trials containing an event with a greater reweighted coherent S/N than the loudest on-source event. Any candidate events with $p$-values below 10% are subjected to additional follow-up studies to determine if the events can be associated with some non-GW noise artifact. Further details on the methods used to search for NS binary signals in coincidence with short GRBs can be found in Harry & Fairhurst (2011) and Williamson et al. (2014).

The efficiency of the NS binary search method for recovering relevant GW signals is evaluated via the addition in software of simulated signals to the data. In order to assess performance, these data are filtered with the same bank of templates used for the search. This provides a means of placing constraints on the short GRB progenitor in the event of no detection in the on-source. All simulated signals are modeled using the "TaylorT2" time-domain, post-Newtonian inspiral approximant (Blanchet et al. 1996). We note that this approximant differs from the one used to build the templates. This choice is designed to account for the disagreement among existing inspiral waveform models in our efficiency assessment (see Nitz et al. 2013 on this topic). Further, this approximant allows for generic spin configurations. We inject three sets of simulated inspiral signals; these correspond to (1) BNS systems with a generic spin configuration, (2) NS-BH systems with a generic spin configuration, and (3) NS-BH systems with an aligned spin configuration. We build both a generic and an aligned spin injection set in the NS-BH case in order to assess the impact of precession on the search sensitivity for rapidly spinning and highly precessing systems (as is the NS-BH case in contrast to the BNS case). The considerations illustrated in Section 3.1 motivate the following choices for the parameters that characterize the three families:

*NS masses:* These are chosen from a Gaussian distribution centered at $1.4\,M_\odot$, with a width of $0.2\,M_\odot$ and $0.4\,M_\odot$ in the BNS and the NS-BH case, respectively (Ozel et al. 2012; Kiziltan et al. 2013). The larger width for NS-BH binaries reflects the greater uncertainty arising from a lack of observed NS-BH systems.

*BH masses:* These are Gaussian distributed with a mean of $10\,M_\odot$ and a width of $6\,M_\odot$. Additionally, they are restricted to being less than $15\,M_\odot$, because the disagreement between different Taylor approximants dominates beyond this point (Nitz et al. 2013).

*Dimensionless spins:* These are drawn uniformly over the intervals [0, 0.4] and [0, 0.98] for NSs and BHs, respectively. For the two sets with generic spin configurations, both spins are isotropically distributed.

[157] A $p$-value is defined as the probability of obtaining such an event or a louder one in the on-source data, given the background distribution, under the null hypothesis.

*Tidal disruption:* NS-BH systems for which the remnant BH is not accompanied by any debris material are not included in the injected populations (Pannarale & Ohme 2014).

*Inclination angle:* This is uniformly distributed in cosine over the intervals [0°, 30°] and [150°, 180°].

*Distance:* Injections are distributed uniformly in distance in the intervals [10, 300] Mpc and [10, 600] Mpc for BNS and NS-BH systems, respectively.

When performing the efficiency assessment, we marginalize over amplitude detector calibration errors by resampling the assumed distance of each injected signal with a Gaussian distribution of 10% width (Abbott et al. 2017b, 2016f); the phase errors of $\sim 5°$ have a negligible effect.

### 4.3. Generic Transient Search Method (X-Pipeline)

Long GRBs are associated with the gravitational collapse of massive stars. While GW emission is expected to accompany such events, its details may vary from event to event. We therefore search for any GW transient without assuming a specific signal shape; this type of search is performed for short GRB events as well. We use the time interval starting from 600 s before each GRB trigger and ending either 60 s after the trigger or at the $T_{90}$ time (whichever is larger) as the *on-source* window to search for a GW signal. This window is large enough to take into account most plausible time delays between a GW signal from a progenitor and the onset of the $\gamma$-ray signal (Koshut et al. 1995; Aloy et al. 2000; MacFadyen et al. 2001; Zhang et al. 2003; Lazzati 2005; Wang & Meszaros 2007; Burlon et al. 2008, 2009; Lazzati et al. 2009; Vedrenne & Atteia 2009). The search is performed on the most sensitive GW band of 16–500 Hz. Above 300 Hz, the GW energy necessary to produce a detectable signal increases sharply as $\propto f^4$ (see Figure 2 of Abbott et al. 2017a), and hence the detector sensitivity is highly biased against high-frequency emission scenarios.

The method used to search for generic GW transients follows the one used in previous GRB analyses (Abadie et al. 2012a; Aasi et al. 2014a, 2014b) and is described in detail in Sutton et al. (2010) and Was et al. (2012). The on-source data for each GRB are processed by the search pipeline to generate multiple time-frequency maps of the data stream using short Fourier transforms with duration at all powers of two between 1/128 s and 2 s. The maps are generated after coherently combining data from the detectors, taking into account the antenna response and noise level of each detector. The time-frequency maps are scanned for clusters of pixels with energy significantly higher than the one expected from background noise. These are referred to as "events" and are characterized by a ranking statistic based on energy. We also perform consistency tests based on the signal correlations measured between the detectors. The event with the highest ranking statistic is taken to be the best candidate for a GW signal for that GRB; it is referred to as the "loudest event." The strategy to associate a $p$-value with the loudest event is the same as the one adopted by the NS binary search but with off-source trials of $\sim$660 s duration.

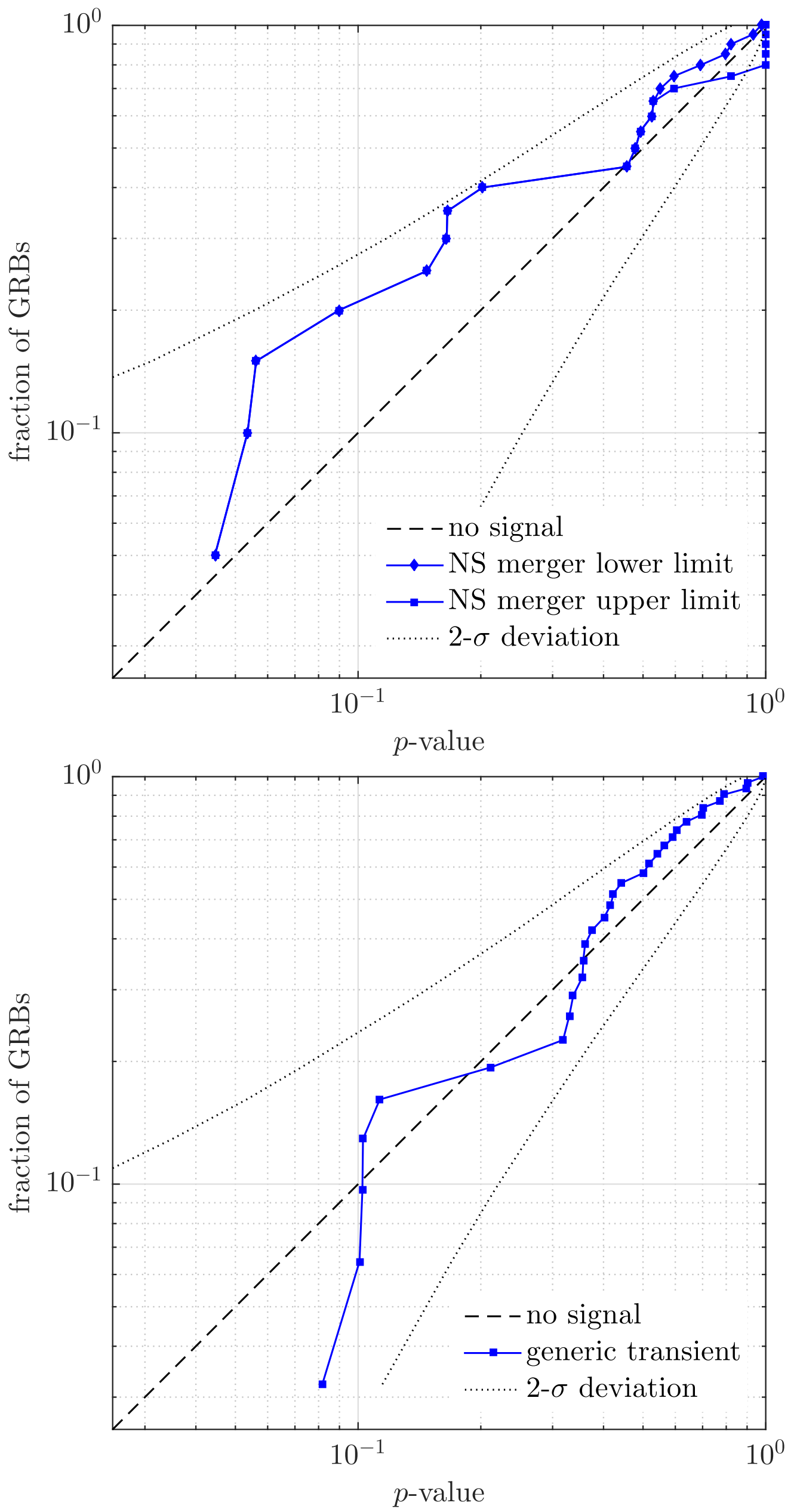

**Figure 2.** Cumulative distribution of $p$-values from the analysis of 20 short-duration GRBs for the evidence of an NS binary merger GW signal (top) and 31 GRBs for the evidence of a GW transient associated with the burst (bottom). The expected distribution under the no-signal hypothesis is indicated by the dashed line, and the $2\sigma$ deviation of that distribution is indicated by the dotted line. For GRBs with no event in the on-source, we provide an upper bound on the $p$-value equal to 1 and a lower bound determined by counting the fraction of background trials that yield no event: this explains the feature in the top right corner of the top panel.

As for the NS binary search method, the efficiency of this search at recovering relevant GW signals is evaluated by the addition in software of simulated signals to the data. The simulated waveforms are chosen to cover the search parameter space; they belong to three types of signals that embrace different potential signal morphologies: NS binary inspiral signals, stellar collapse (represented by CSGs), and disk instability models (represented by ADI

**Table 2**
Median 90% Confidence Level Exclusion Distances $D_{90\%}$

| Short GRBs | BNS | | NS-BH Aligned Spins | | NS-BH Generic Spins |
|---|---|---|---|---|---|
| $D_{90\%}$ [Mpc] | 90 | | 150 | | 139 |
| All GRBs | CSG 70 Hz | CSG 100 Hz | | CSG 150 Hz | CSG 300 Hz |
| $D_{90\%}$ [Mpc] | 88 | 89 | | 71 | 30 |
| All GRBs | ADI A | ADI B | ADI C | ADI D | ADI E |
| $D_{90\%}$ [Mpc] | 31 | 97 | 39 | 15 | 36 |

**Notes.** The short GRB analysis assumes an NS binary progenitor. When all GRBs are analyzed, a circular sine-Gaussian (CSG) or an accretion disk instability (ADI) model is used.

waveforms).[158] In particular, the generic time-frequency excess power method used is equally efficient for descending (ADI) and ascending (NS binary) chirps. Because this paper reports results for NS binaries only when these are obtained with the dedicated, modeled search outlined in Section 4.2, we will limit the discussion to the case of the other two signal families.

*CSG:* For the standard siren CSG signals defined in Equation (1), we assume an optimistic emission of energy in GWs of $E_{GW} = 10^{-2}\, M_\odot c^2$. As discussed in Section 3, this is an upper bound on the predictions: our conclusions thus represent upper bounds, as we work under the optimistic assumption that every GRB emits $10^{-2}\, M_\odot c^2$ of energy in GWs. Further, we construct four sets of such waveforms with a fixed $Q$ factor of 9 and varying center frequency (70, 100, 150, and 300 Hz).

*ADI:* The extreme scenario of ADIs (van Putten 2001; van Putten et al. 2004) provides long-lasting waveforms that the unmodeled search has the ability to recover. We chose the same sets of parameters used in a previous long-transient search (Abbott et al. 2016b) to cover the different predicted morphologies. The values of the parameters are listed in Table 1. As in previous searches, the clumps in the disk are assumed to be forming at a distance of 100 km from the BH innermost stable circular orbit (Ott & Santamaría 2013), which is the typical distance of the transition to a neutrino opaque disk where the accretion disk is expected to have the largest linear density (Lee et al. 2005; Chen & Beloborodov 2007). This constitutes a deviation from the original model that brings the GW emission from ∼1 kHz to a few hundred Hz, where the detectors are more sensitive, thus providing a reasonable means of testing the ability of the search to detect signals in this frequency band and with amplitudes comparable to the original ADI formulation. We note that in the previous search for long-duration signals associated with GRBs (Aasi et al. 2013), these signals were normalized to obtain $E_{GW} = 0.1\, M_\odot c^2$. These waveforms are tapered by a Tukey window with 1 s at the start and end of the waveform to avoid artifacts from the unphysical sharp start and end of these waveforms.

[158] In general, the sensitivity of an excess power search compared to ideal match filtering scales as $V^{0.25}$, where $V$ is the time-frequency volume of the signal (Anderson et al. 2001). In practice, the sensitivity of this search compared to ideal match filtering is similar for CSGs, a factor of ∼2 poorer for NS merger signals, and a factor of ∼3 poorer for ADI signals. We refer the interested reader to Sutton et al. (2010) for further details.

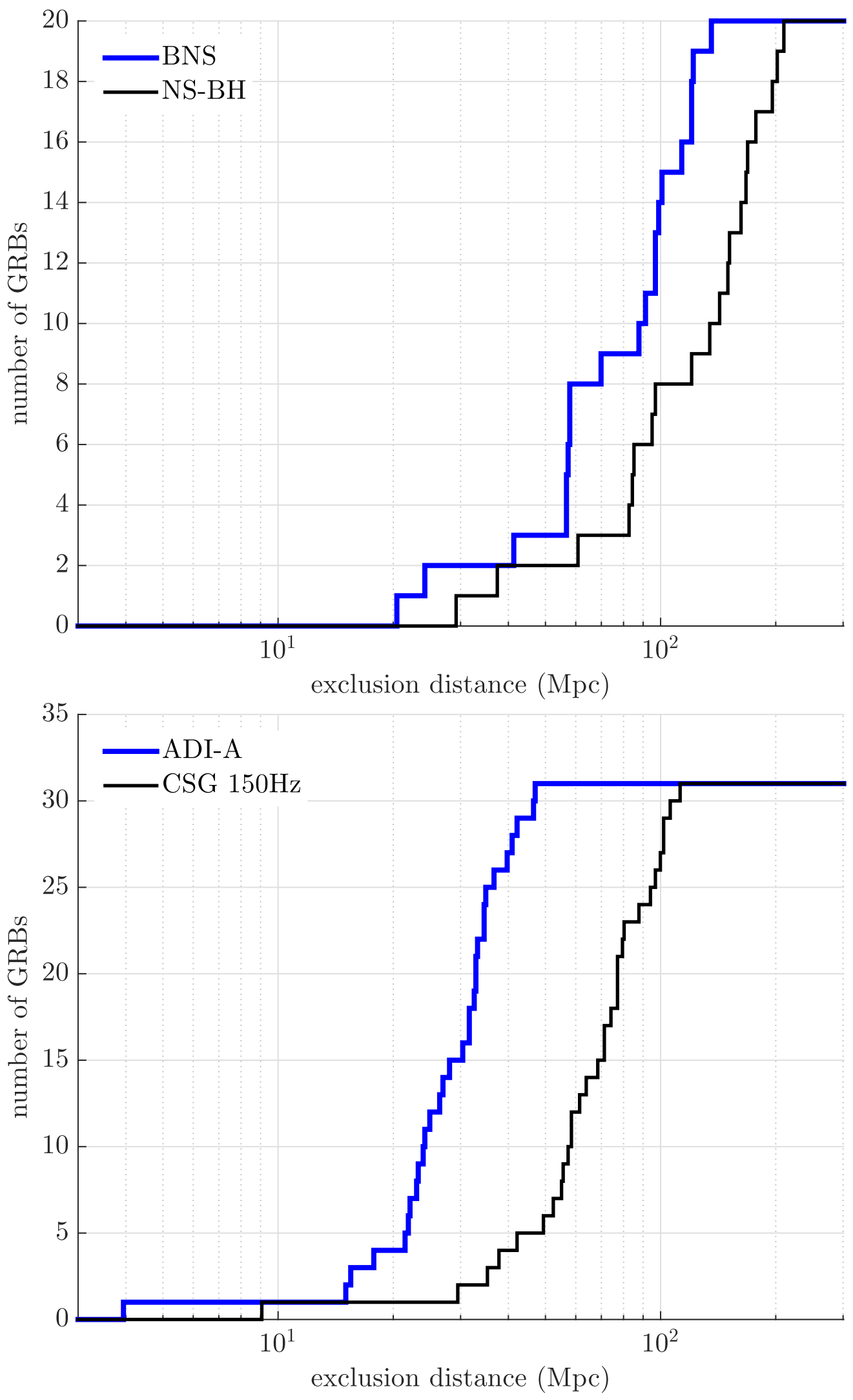

**Figure 3.** Cumulative histograms of the exclusion distances at the 90% confidence level for BNS and NS-BH systems across the sample of short GRBs (top) and for ADI-A and CSG GW transients at 150 Hz across the sample of all GRBs analyzed with the generic transient search (bottom). Both ADI-A and CSG at 150 Hz signals have an emission energy $\sim 10^{-2} M_\odot c^2$, but for ADI-A the energy is spread over a ∼100 times longer duration, which explains the difference in exclusion distances.

Finally, calibration errors are folded into the result by jittering the signal amplitude and time of arrival at each detector, following a wider Gaussian distribution of 20% in amplitude and 20 degrees in phase, as this search used the preliminary Advanced LIGO calibration that had greater uncertainties (Tuyenbayev et al. 2017).

## 5. Results

A search for GWs in coincidence with GRBs was performed during O1. We analyzed a total of 31 GRBs using the generic transient method and 19 GRBs, classified as short or ambiguous, using the NS binary search method. In addition, we used the NS binary search method to analyze GRB 150906B, which occurred prior to 2015 September 12. The detailed list of analyzed GRBs and the search results are provided in Table 3 in the Appendix.

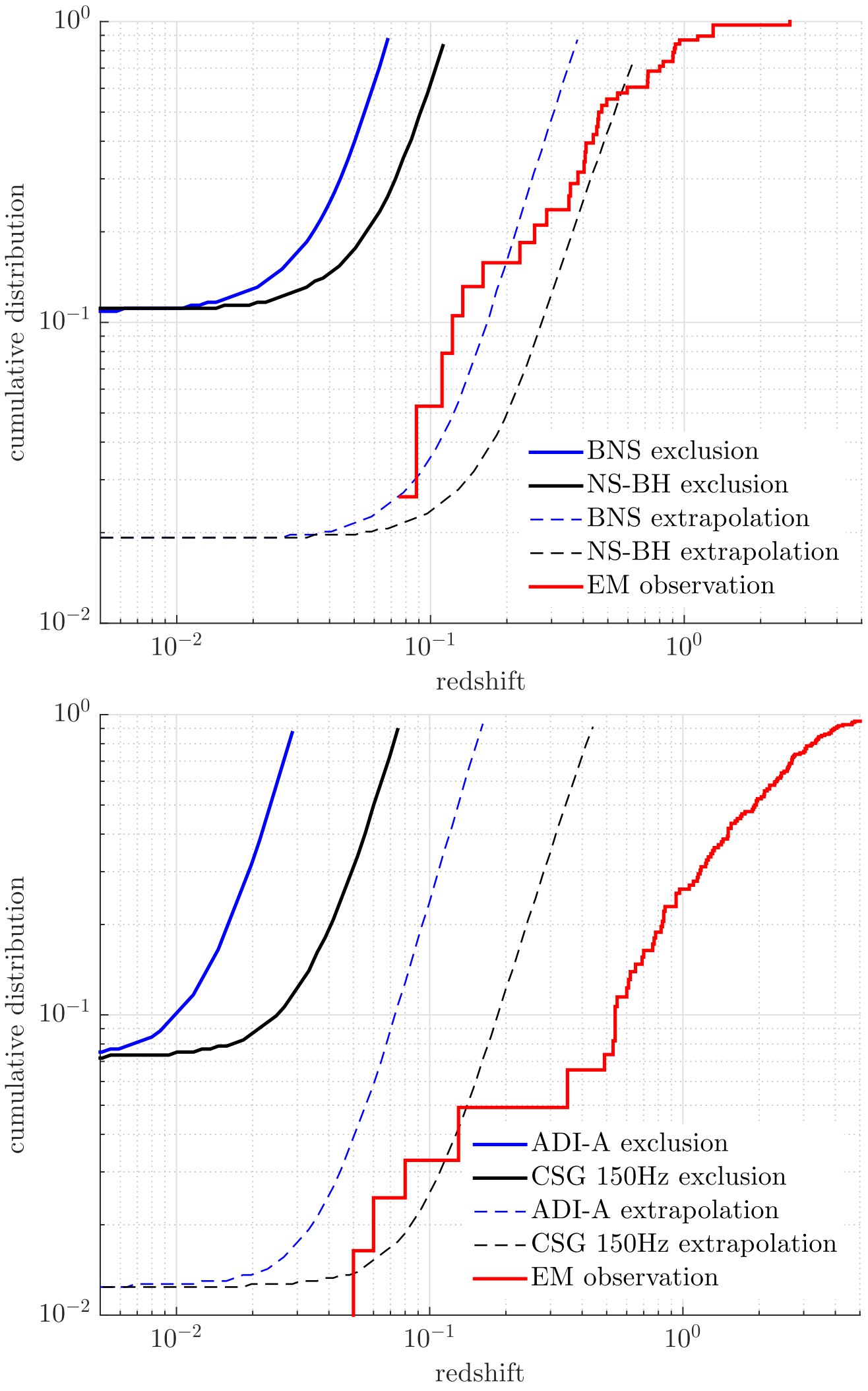

**Figure 4.** Combined exclusion distance for 20 short GRBs analyzed with the coalescence search for both a BNS and an NS-BH progenitor (top) and for all 31 GRBs analyzed with the generic transient search for ADI-A and standard siren CSG GW transients at 150 Hz with an energy of $E_{\rm GW} = 10^{-2} M_{\odot} c^2$ (bottom). We exclude at 90% confidence level cumulative distance distributions that pass through the region above the solid curves. For reference, the red staircase curve shows the cumulative distribution of measured redshifts for short GRBs (top; Leibler & Berger 2010; Fong et al. 2015; Siellez et al. 2016) and *Swift* GRBs (bottom; Jakobsson et al. 2006, 2012). The dashed curves are an extrapolation of these results to 2 years of Advanced LIGO operation at design sensitivity.

Overall, the RAVEN (Urban 2016) analysis yielded no temporal coincidences between GW candidates from low-latency searches and GRB triggers. With the two offline searches, we found no noteworthy individual events, nor evidence for a collective signature of weak GW signals associated with the GRB population. The distribution of observed $p$-values is shown in Figure 2; for GRBs with no event in the on-source, we provide an upper bound on the $p$-value equal to 1 and a lower bound determined by counting the fraction of background trials that yield no event: this explains the feature in the top right corner of the top panel. These $p$-values are combined using the weighted binomial test (Abadie et al. 2012a) to quantitatively assess the population consistency with the no-signal hypothesis. This test looks at the lowest 5% of $p$-values weighted by the prior probability of detection based on the GW sensitivity at the time of and from the direction of the GRB. The NS binary (generic transient) search method yielded a combined $p$-value of 57% (75%).

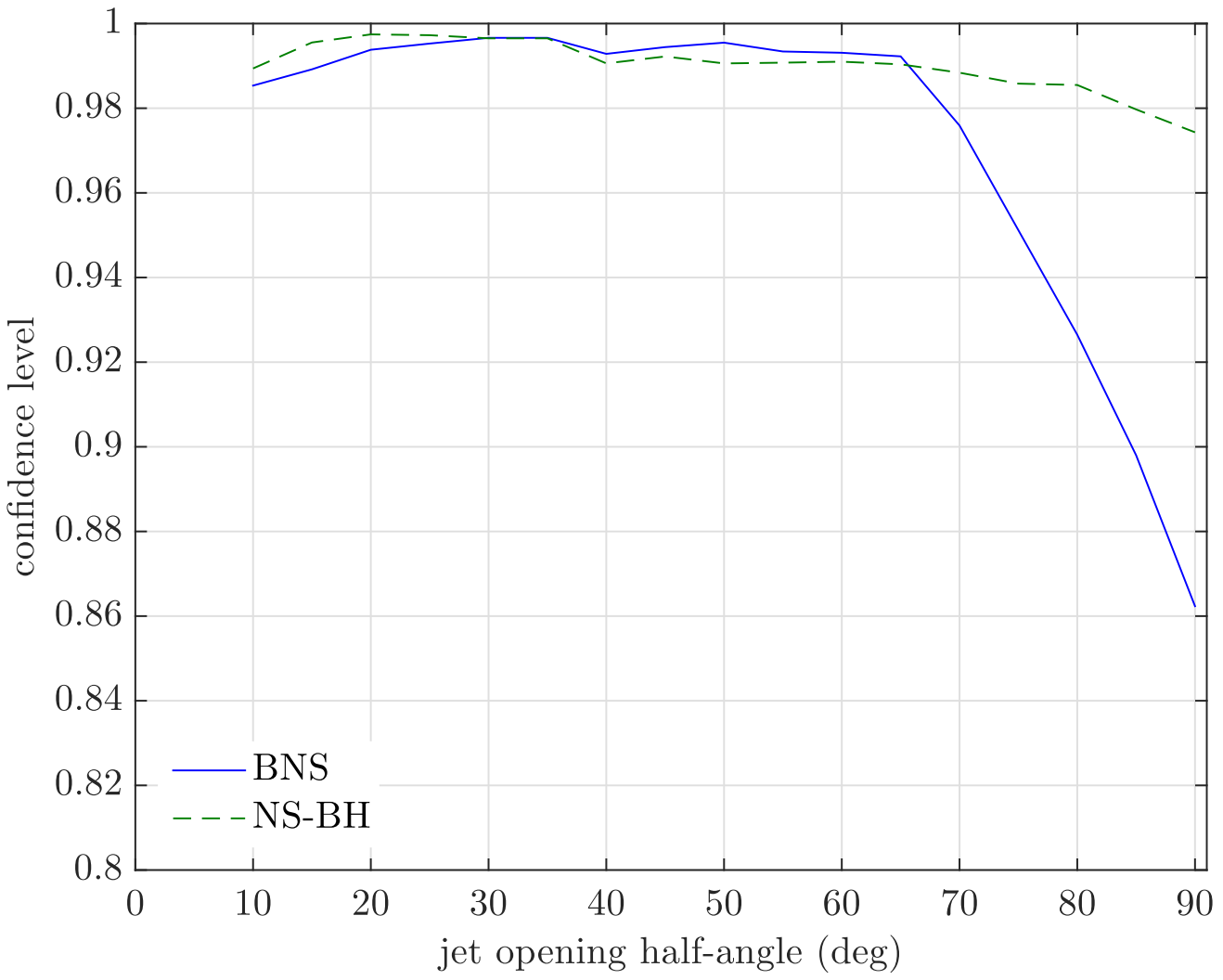

**Figure 5.** Exclusion confidence level for binaries at 54 Mpc from Earth as a function of the jet opening half-angle $\theta_{\rm jet}$ of the binary. The simulated signals were performed with a uniform distribution in the cosine of the inclination angle $\iota$, hence with a small number of cases at low $\iota$. This causes a small decrease in confidence for jet angles below 20° due to a larger statistical uncertainty.

Given that the analyses returned no significant event, we place limits on GW emission based both on binary mergers in the case of short GRBs and on generic GW transient signal models for all 42 GRBs in our sample. For a given signal morphology, the GW analysis efficiently recovers signals up to a certain distance that depends on the sensitivity of the detectors at the time and sky position of a given GRB event. We quote a 90% confidence level lower limit on the distance $D_{90\%}$ to each GRB progenitor, that is, the distance at which 90% of simulated signals are recovered with a ranking statistic that is greater than the largest value actually measured. The quoted exclusion distances are marginalized over systematic errors introduced by the mismatch of a true GW signal and the waveforms used in the simulations, and over amplitude and phase errors from the calibration of the detector data. The median exclusion distances are summarized in Table 2, while the cumulative distributions of exclusion distances for a subset of injected signal populations are shown in Figure 3. For short GRBs, the median exclusion distance is between 90 and 150 Mpc depending on the assumed NS binary progenitor, whereas for all GRBs and a generic GW signal model, the median exclusion distance is between 15 Mpc and

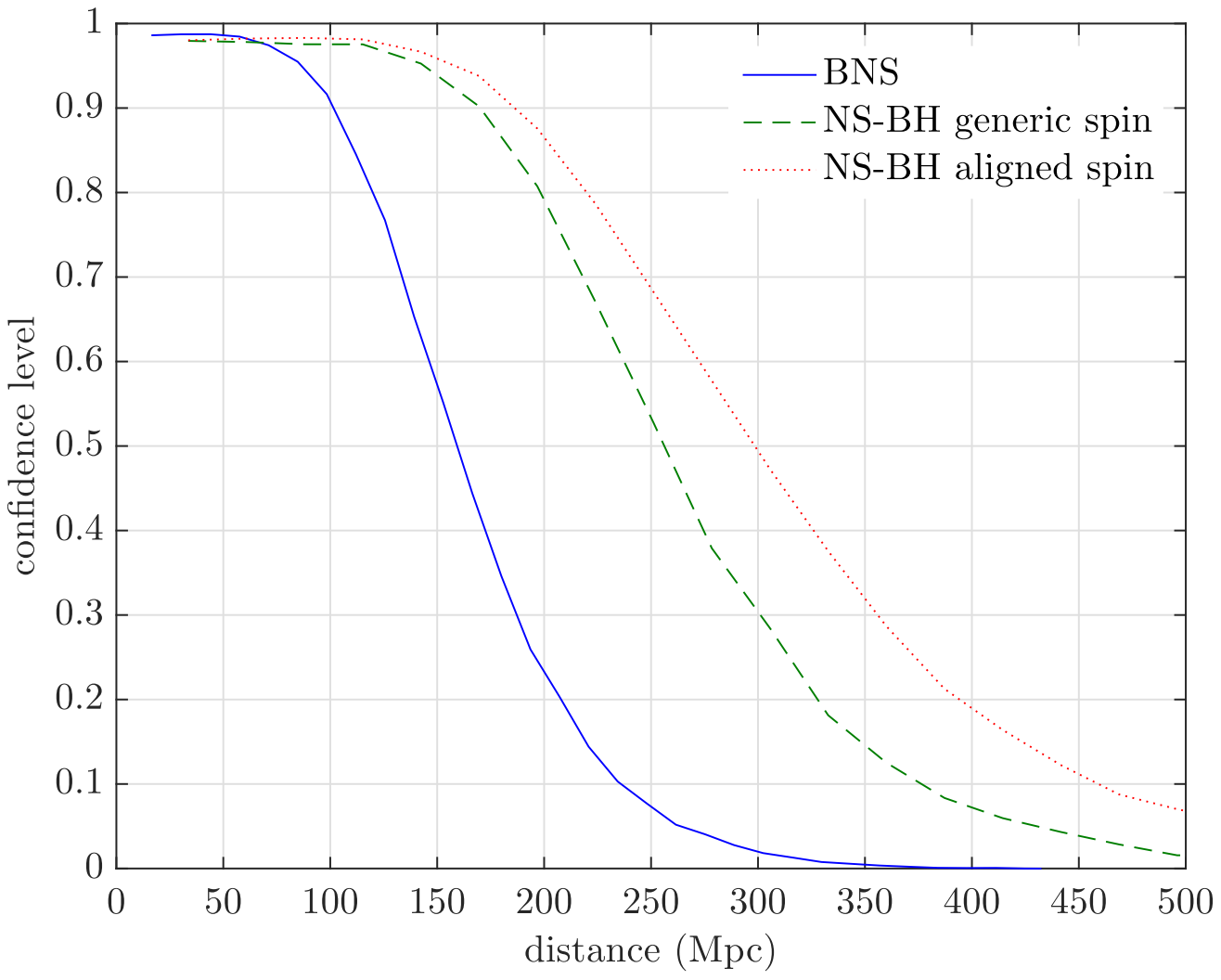

**Figure 6.** Exclusion confidence level for three populations of simulated binary merger signals as a function of distance, given LIGO observations at the time of GRB 150906B.

100 Mpc. The results for the NS binary search can be compared to the ranges reported in Tables 1 and 2 of Abbott et al. (2016c) for the all-time, all-sky search for GWs emitted by BNS and NS-BH systems in O1. Both searches are most sensitive to aligned-spin NS-BH binaries and least sensitive to BNS binaries. This hierarchy is determined by the masses and by the degree of spin misalignment involved in the simulated source populations: all else being equal, GW detectors are less sensitive to lower mass systems because these have smaller GW amplitudes, while searches performed with aligned spin templates progressively lose in efficiency as precession effects in the source become more and more marked. Further, as discussed by Williamson et al. (2014), the targeted, coherent search is sensitive to distances that are 20%–25% greater than those achieved by a coincident all-sky search. This explains why the distances reported here are greater than those in Abbott et al. (2016c). Clearly, this is a rough comparison because the injected populations considered here and by the all-sky all-time search are different, particularly with regards to the choice of BH masses and to the restriction set on the inclination angle.

By combining results from all analyzed GRBs, we place exclusions on GRB progenitor populations. To do this, we use a simple population model, where all GRB progenitors have the same GW emission (standard sirens), and perform exclusions on cumulative distance distributions. We parameterize the distance distribution with two components: a fraction $F$ of GRBs distributed with a constant comoving density rate up to a luminosity distance $R$, and a fraction $1-F$ at effectively infinite distance. This simple model yields a parameterization of astrophysical GRB distance distribution models that predict a uniform local rate density and a more complex dependence at redshift higher than 0.1, given that the high-redshift part of the distribution is beyond the sensitivity of current GW detectors. The exclusion is then performed in the $(F, R)$ plane. (For details of this method, see Appendix B of Abadie et al. 2012a.) The exclusion for BNS and NS-BH sources is shown in the top panel of Figure 4. The bottom panel instead shows the exclusion for the ADI-A model and for GW transient signals modeled as CSGs at 150 Hz, under the optimistic assumption that the energy emitted in GWs by every GRB is $E_{\rm GW} = 10^{-2} M_\odot c^2$. For comparison, we plot the redshift distribution of short GRBs (or for all GRBs observed by *Swift*). In neither case does the exclusion line come close to the observed population redshift, indicating that it would have been unlikely to observe an event in this analysis.

An extrapolation of these results to 2 years of operation at Advanced LIGO design sensitivity, which is a factor of $\sim$3 better than the one obtained during O1 (Abbott et al. 2016a; Martynov et al. 2016), is shown in Figure 4. For short GRBs, the observations will then probe the nearby tail of the distribution and therefore the validity of the NS binary merger origin of short GRBs. Long GRB observations, however, will only probe nearby faint GRB events at redshift $\sim$0.1, either achieving a detection from a nearby GRB or excluding that all nearby long GRBs have a very energetic GW emission with $E_{\rm GW} \sim 10^{-2} M_\odot c^2$. In this respect, under the less optimistic assumption that $E_{\rm GW} \sim 10^{-4} M_\odot c^2$ for all nearby long GRBs, would shift the extrapolated CSG exclusion region to redshifts that are an order of magnitude lower (see, e.g., Figure 7 in Aasi et al. 2014b). These extrapolations and conclusions are consistent with previous extrapolations (Aasi et al. 2014b).

### 5.1. GRB 150906B

If NGC 3313 were indeed the host of an NS binary merger progenitor of GRB 150906B, Advanced LIGO should have detected a GW signal associated with the event, given the proximity of this galaxy located at a luminosity distance of 54 Mpc from Earth. A similar hypothesis was previously tested with the initial LIGO detectors for GRB 051103 and GRB 070201, the error boxes of which overlapped the M81/M82 group of galaxies and M31, respectively (Abbott et al. 2008; Abadie et al. 2012b). In both cases, a binary merger scenario was excluded with greater than 90% confidence, and the preferred scenario is that these events were extragalactic soft-gamma-repeater flares.

The NS binary search described in Section 4.2 found no evidence for a GW signal produced at the time and sky position of GRB 150906B. The most significant candidate event in the on-source region around the time of the GRB had a $p$-value of 53%.

This null-detection result allows us to compute the frequentist confidence with which our search excludes a binary coalescence in NGC 3313. This confidence includes both the search efficiency at recovering signals as well as our uncertainty in measuring such efficiency. Figure 5 shows the exclusion confidence for BNS and NS-BH systems as a function of the jet half-opening angle $\theta_{\rm jet}$, assuming a distance[159] to NGC 3313 of 54 Mpc and that the NS binary inclination angle $\iota$ between the total angular momentum axis and the line of sight is distributed uniformly in $\cos\iota$ up to $\theta_{\rm jet}$. If we assume an isotropic (i.e., unbeamed) $\gamma$-ray emission from

[159] To account for the detector calibration errors in the pre-O1 stage during which GRB 150906B occurred, the simulated signals added in software to the data for this study were jittered with a Gaussian distribution of 20% in amplitude and 20° in phase.

GRB 150906B, the possibility of a BNS coalescence progenitor is excluded with $\gtrsim 86\%$ confidence. Taking a fiducial jet half-opening angle upper limit of 30° (or equivalently a maximum binary inclination angle of this size), the exclusion confidence rises to $\gtrsim 99.7\%$. NS-BH systems with isotropic emission are excluded at $\gtrsim 97\%$ confidence, which rises to $\gtrsim 99.7\%$ for $\theta_{\rm jet} \leqslant 30°$.

The increase in exclusion confidence for smaller jet angles is due to the fact that the average amplitude of the GW signal from an NS binary coalescence is larger for systems for which the orbital plane is viewed "face-on" (where the detector receives the flux from both GW polarizations) than for systems viewed "edge-on" (where the detector receives the flux from just one GW polarization); small jet angles imply a system closer to face-on.

To determine the distance up to which we can exclude, with 90% confidence, a binary coalescence as the progenitor of GRB 150906B, we assume beamed emission with a maximum opening angle of 30° and compute the distance at which 90% of injected BNS, generic spin NS-BH, and aligned spin NS-BH signals are recovered louder than the loudest on-source event. The result is shown in Figure 6. BNS systems are excluded with 90% confidence out to a distance of 102 Mpc, while generic and aligned spin NS-BH systems are excluded with the same confidence at 170 Mpc and 186 Mpc, respectively. This is consistent with theoretical arguments based on $\gamma$-ray spectrum and fluence that place the progenitor of GRB 150906B at more than 270 Mpc (Ruffini et al. 2015; Zhang et al. 2015), possibly in one of the several known galaxies at around 500 Mpc within the error region (Bilicki et al. 2013).

## 6. Conclusion

We have analyzed data from O1 to look for GWs coincident with GRBs that occurred during this period, using both a modeled search for BNS and NS-BH systems and an unmodeled search for GW transients. No GW was detected in coincidence with a GRB by either search. We set lower limits on the distance for each GRB for various GW emission models. The median of these distance limits is higher than distance limits placed by all previous modeled and unmodeled searches (e.g., Abadie et al. 2012a; Aasi et al. 2014b). We also combined these lower limits into an exclusion on the GRB redshift distribution. This exclusion is a factor of a few away from the short and long GRB distributions measured by $\gamma$-ray satellites.

With 2 years of observation at design sensitivity, Advanced LIGO will probe the observed redshift distribution. At that point, either a GW detection in association with a short GRB will take place, or the result will be in tension with the NS binary merger progenitor scenario for short GRBs. For long GRBs, a lack of detection would only constrain the most extreme scenarios of GW emission from a strongly rotating stellar core collapse.

We also analyzed data from the LIGO Hanford detector to look for a GW signal associated with GRB 150906B. No evidence was found for a GW signal associated with this GRB. The sensitivity of the modeled search allows us to confidently exclude the hypothesis that an NS binary in NGC 3313 was the progenitor of GRB 150906B. If the event indeed occurred in NGC 3313, then it would have had to defy the setup of the modeled search. In this case, and in light of the problematic classification of GRB 150906B discussed in Section 2, this GRB may most probably have been due to a stellar core collapse or a giant flare from a soft-gamma repeater. Alternatively, GRB 150906B may have simply originated from an NS binary merger in one of the more distant galaxies at 500 Mpc, compatible with the sky location of the event.

The authors gratefully acknowledge the support of the United States National Science Foundation (NSF) for the construction and operation of the LIGO Laboratory and Advanced LIGO as well as the Science and Technology Facilities Council (STFC) of the United Kingdom, the Max-Planck-Society (MPS), and the State of Niedersachsen/Germany for support of the construction of Advanced LIGO and construction and operation of the GEO600 detector. Additional support for Advanced LIGO was provided by the Australian Research Council. The authors gratefully acknowledge the Italian Istituto Nazionale di Fisica Nucleare (INFN), the French Centre National de la Recherche Scientifique (CNRS), and the Foundation for Fundamental Research on Matter supported by the Netherlands Organisation for Scientific Research for the construction and operation of the Virgo detector and the creation and support of the EGO consortium. The authors also gratefully acknowledge research support from these agencies as well as by the Council of Scientific and Industrial Research of India, Department of Science and Technology, India, Science & Engineering Research Board (SERB), India, Ministry of Human Resource Development, India, the Spanish Ministerio de Economía y Competitividad, the Conselleria d'Economia i Competitivitat and Conselleria d'Educació Cultura i Universitats of the Govern de les Illes Balears, the National Science Centre of Poland, the European Commission, the Royal Society, the Scottish Funding Council, the Scottish Universities Physics Alliance, the Hungarian Scientific Research Fund (OTKA), the Lyon Institute of Origins (LIO), the National Research Foundation of Korea, Industry Canada and the Province of Ontario through the Ministry of Economic Development and Innovation, the Natural Science and Engineering Research Council Canada, Canadian Institute for Advanced Research, the Brazilian Ministry of Science, Technology, and Innovation, Fundação de Amparo à Pesquisa do Estado de São Paulo (FAPESP), Russian Foundation for Basic Research, the Leverhulme Trust, the Research Corporation, Ministry of Science and Technology (MOST), Taiwan and the Kavli Foundation. The authors gratefully acknowledge the support of the NSF, STFC, MPS, INFN, CNRS, and the State of Niedersachsen/Germany for provision of computational resources. K. Hurley is grateful for IPN support under NASA grant NNX15AU74G. R.L. Aptekar and S.V. Golenetskii gratefully acknowledge support from RFBR grants 15-02-00532-i and 16-29-13009-ofi-m.

## Appendix
## Results Table

**Table 3**
Analyzed GRB Sample and Search Results

| GRB Name | UTC Time | R.A. | Decl. | Satellite(s) | Type | Network | $D_{90\%}$ (Mpc) | | | | |
|---|---|---|---|---|---|---|---|---|---|---|---|
| | | | | | | | | NS-BH | | GW Burst | |
| | | | | | | | BNS | Generic Spins | Aligned Spins | ADI A | CSG 150 Hz |
| 150906B | 08:42:25 | $10^{h}36^{m}57^{s}$ | −25°36′ | IPN | Ambiguous | H1 | 102 | 170 | 186 | ⋯ | ⋯ |
| 150912600 | 14:24:31 | $21^{h}25^{m}26^{s}$ | 73°16′ | *Fermi* | Short | H1L1 | 88 | 150 | 150 | 28 | 71 |
| 150912A | 10:37:38 | $16^{h}33^{m}46^{s}$ | −21°02′ | *Fermi* | Long | H1L1 | ⋯ | ⋯ | ⋯ | 47 | 113 |
| 150919A | 20:43:18 | $08^{h}51^{m}50^{s}$ | 44°04′ | IPN | Short | H1 | 58 | 83 | 102 | ⋯ | ⋯ |
| 150922883 | 21:11:32 | $18^{h}16^{m}34^{s}$ | −50°28′ | *Fermi* | Ambiguous | H1L1 | 71 | 122 | 122 | ⋯ | ⋯ |
| 150922A | 05:37:29 | $19^{h}31^{m}50^{s}$ | −2°15′ | *Fermi* | Short | H1L1 | 100 | 163 | 183 | 27 | 69 |
| 150923297 | 07:07:36 | $21^{h}07^{m}12^{s}$ | 31°49′ | *Fermi* | Short | H1L1 | 98 | 144 | 187 | 33 | 88 |
| 150923429 | 10:18:17 | $17^{h}51^{m}14^{s}$ | −40°40′ | *Fermi* | Short | H1L1 | 136 | 213 | 241 | 41 | 97 |
| 150925A | 04:09:28 | $15^{h}10^{m}08^{s}$ | −19°38′ | *Swift* | Long | H1L1$^{\dagger}$ | ⋯ | ⋯ | ⋯ | 23 | 50 |
| 151001348 | 08:20:35 | $16^{h}26^{m}57^{s}$ | −10°08′ | *Fermi* | Long | H1L1$^{\dagger}$ | ⋯ | ⋯ | ⋯ | 33 | 81 |
| 151006A | 09:55:01 | $9^{h}49^{m}42^{s}$ | 70°30′ | *Swift* | Long | H1L1$^{\dagger}$ | ⋯ | ⋯ | ⋯ | 31 | 64 |
| 151009949 | 22:47:03 | $14^{h}48^{m}00^{s}$ | 63°43′ | *Fermi* | Long | H1L1 | ⋯ | ⋯ | ⋯ | 42 | 95 |
| 151019 | 08:05:28 | $6^{h}37^{m}49^{s}$ | 79°08′ | IPN | Long | H1L1 | ⋯ | ⋯ | ⋯ | 15 | 30 |
| 151022577 | 13:51:02 | $7^{h}21^{m}28^{s}$ | 40°14′ | *Fermi* | Short | H1L1 | 115 | 179 | 206 | 47 | 102 |
| 151022A | 14:06:32 | $23^{h}16^{m}47^{s}$ | 55°49′ | *Swift* | Long | H1L1$^{\dagger}$ | ⋯ | ⋯ | ⋯ | 25 | 58 |
| 151023A | 13:43:04 | $18^{h}03^{m}56^{s}$ | −8°19′ | *Swift* | Long | H1L1 | ⋯ | ⋯ | ⋯ | 35 | 80 |
| 151024179 | 04:17:53 | $15^{h}31^{m}26^{s}$ | 22°57′ | *Fermi* | Ambiguous | H1 | 25 | 30 | 48 | ⋯ | ⋯ |
| 151027B | 22:40:40 | $5^{h}04^{m}52^{s}$ | −6°27′ | *Swift* | Long | H1L1$^{\dagger}$ | ⋯ | ⋯ | ⋯ | 40 | 102 |
| 151029A | 07:49:39 | $2^{h}34^{m}08^{s}$ | −35°21′ | *Swift* | Long | H1L1 | ⋯ | ⋯ | ⋯ | 16 | 35 |
| 151107B | 20:24:52 | $2^{h}05^{m}12^{s}$ | 45°35′ | *Fermi* | Long | H1L1$^{\dagger}$ | ⋯ | ⋯ | ⋯ | 4 | 9 |
| 151112A | 13:44:48 | $0^{h}08^{m}12^{s}$ | −61°40′ | *Swift* | Long | H1L1 | ⋯ | ⋯ | ⋯ | 37 | 100 |
| 151114A | 09:59:50 | $8^{h}03^{m}45^{s}$ | −61°03′ | *Swift* | Ambiguous | L1 | 42 | 61 | 75 | ⋯ | ⋯ |
| 151117 | 01:37:03 | $1^{h}44^{m}32^{s}$ | 18°39′ | IPN | Long | H1L1 | ⋯ | ⋯ | ⋯ | 33 | 77 |
| 151121 | 06:56:27 | $19^{h}35^{m}22^{s}$ | 7°20′ | IPN | Short | H1L1 | ⋯ | ⋯ | ⋯ | 32 | 59 |
| 151126 | 04:03:03 | $13^{h}05^{m}20^{s}$ | 0°07′ | IPN | Short | H1L1 | 122 | 203 | 217 | 35 | 78 |
| 151127A | 09:08:49 | $1^{h}17^{m}54^{s}$ | −82°46′ | *Swift* | Short | H1L1 | 97 | 152 | 165 | 33 | 78 |
| 151130160 | 03:50:50 | $9^{h}05^{m}04^{s}$ | −18°49′ | *Fermi* | Long | H1L1 | ⋯ | ⋯ | ⋯ | 18 | 62 |
| 151202565 | 13:33:49 | $21^{h}45^{m}58^{s}$ | −24°40′ | *Fermi* | Short | H1 | 121 | 198 | 226 | ⋯ | ⋯ |
| 151218857 | 20:33:31 | $0^{h}37^{m}48^{s}$ | −30°44′ | *Fermi* | Ambiguous | H1L1 | 21 | 38 | 35 | ⋯ | ⋯ |
| 151219 | 09:11:16 | $14^{h}34^{m}01^{s}$ | 12°57′ | IPN | Long | H1L1 | ⋯ | ⋯ | ⋯ | 24 | 55 |
| 151219567 | 13:36:22 | $23^{h}24^{m}45^{s}$ | 11°22′ | *Fermi* | Long | H1L1$^{\dagger}$ | ⋯ | ⋯ | ⋯ | 32 | 71 |
| 151222A | 08:10:13 | $23^{h}40^{m}43^{s}$ | 36°42′ | *Fermi* | Short | H1L1 | 59 | 96 | 104 | 22 | 38 |
| 151227A | 01:44:07 | $13^{h}42^{m}00^{s}$ | 65°52′ | *Fermi* | Ambiguous | H1L1 | 57 | 97 | 108 | 23 | 56 |
| 151227B | 05:13:48 | $19^{h}11^{m}33^{s}$ | 31°56′ | *Fermi* | Long | H1L1 | ⋯ | ⋯ | ⋯ | 22 | 53 |
| 151228A | 03:05:12 | $14^{h}16^{m}01^{s}$ | −17°41′ | *Swift* | Short | H1 | 122 | 169 | 200 | ⋯ | ⋯ |
| 151229486 | 11:40:06 | $23^{h}05^{m}58^{s}$ | 6°55′ | *Fermi* | Short | H1 | 57 | 86 | 93 | ⋯ | ⋯ |
| 151231A | 10:37:47 | $4^{h}22^{m}31^{s}$ | −61°32′ | *Fermi* | Long | H1L1$^{\dagger}$ | ⋯ | ⋯ | ⋯ | 27 | 75 |
| 151231B | 13:38:08 | $10^{h}00^{m}19^{s}$ | 28°49′ | *Fermi* | Short | L1 | 58 | 85 | 96 | ⋯ | ⋯ |
| 160101A | 00:43:53 | $14^{h}38^{m}36^{s}$ | −13°49′ | *Fermi* | Long | H1L1 | ⋯ | ⋯ | ⋯ | 35 | 107 |
| 160103 | 17:39:04 | $13^{h}14^{m}53^{s}$ | −23°26′ | IPN | Long | H1L1 | ⋯ | ⋯ | ⋯ | 22 | 42 |
| 160111115 | 02:45:03 | $20^{h}40^{m}57^{s}$ | −32°47′ | *Fermi* | Long | H1L1 | ⋯ | ⋯ | ⋯ | 24 | 59 |
| 160111A | 07:22:02 | $03^{h}02^{m}31^{s}$ | 28°51′ | IPN | Short | H1 | 91 | 135 | 151 | ⋯ | ⋯ |

**Notes.** Information and limits on associated GW emission for each of the analyzed GRBs. The first six columns are as follows: the GRB name in YYMMDD format, the trigger time, the sky position used for the GW search (R.A. and decl.), the satellite whose sky localization is used, and the GRB classification type. The seventh column gives the GW detector network used: here H1 refers to the interferometer in Hanford, WA, and L1 to the one in Livingston, LA; a $^{\dagger}$ denotes cases in which the on-source window of the generic transient search is extended to cover the GRB duration ($T_{90} > 60$ s). Columns 8–12 display the 90% confidence lower limits on the exclusion distance to the GRB ($D_{90\%}$) for several emission scenarios: BNS, generic and aligned spin NS-BH, accretion disk instability (ADI)-A, and circular sine-Gaussian (CSG) GW burst at 150 Hz with total radiated energy $E_{\rm GW} = 10^{-2} M_{\odot} c^{2}$. When the use of only the generic transient or the NS binary search method was possible, only a subset of exclusion distances is shown. For GRB 150922883 and GRB 151218857, there were not enough data from both LIGO detectors to run the generic GW transient search, so results are reported for the NS binary coalescence search only. The short GRB 151121 was localized by the IPN with an error box area of about 106 square degrees; it was therefore not analyzed with the modeled search due to the high computational costs this would have required and the negligible increase in sensitivity rendered by a targeted search (Aasi et al. 2014b).